\documentclass[prd,tightenlines,nofootinbib,showpacs]{revtex4}
\def\journal#1, #2, #3#4, #5#6#7#8    {
    {#1~}{#2} (#5#6#7#8) #3#4}

\def\prb{\journal Phys. Rev. B, }

\def\prl{\journal Phys. Rev. Lett., }

\def\cmp{\journal Comm. Math. Phys., }

\def\npb{\journal Nucl. Phys. B, }
\def\plb{\journal Phys. Lett. B, }

\def\mpla{\journal Mod. Phys. Lett. A, }
\def\ijmpa{\journal Int. Jour. Mod. Phys. A, }
\def\jpsj{\journal J. Phys. Soc. Japan, }
\def\jmp{\journal J. Math. Phys., }

\def\jhep{\journal J. High Energy Phys., }
\def\Rb{{I\!\! R}}
\def\Nb{{I\!\! N}}

\def\JM{{\bf 1}\hspace{-0.12cm}{\rm I}}

\newcommand{\beq}[1]{\begin{equation}\label{#1}}
\newcommand\eeq{\end{equation}}
\newcommand{\ba}[1]{\begin{eqnarray}\label{#1}}
\newcommand{\baa}{\begin{eqnarray}}
\newcommand\ea{\end{eqnarray}}
\newcommand{\bee}{\begin{equation}}
\newcommand{\br}[1]{\overline #1}
\def\nn{\nonumber \\}
\def\l{\lambda}
\def\n{\nu}

\def\A{\tilde A}

\def\Tr{{\rm Tr}}
\newcommand{\h}{Hamiltonian}

\def\hlf{\frac{1}{2}}

\newcommand{\va}{|0,l\rangle}

\newcommand{\vac}{|0\rangle}

\begin{document}

\title{Algebra of observables in the Calogero model and in the 
Chern-Simons matrix model}

\author{Larisa Jonke}
\email{larisa@irb.hr}
\author{Stjepan Meljanac}
\email{meljanac@irb.hr}
\affiliation{Theoretical Physics Division,\\
Rudjer Bo\v skovi\'c Institute, P.O. Box 180,\\
HR-10002 Zagreb, Croatia}

\begin{abstract}
The algebra of observables of an $N$-body Calogero model is
represented on the $S_N$-symmetric subspace of the positive definite 
Fock space.   We discuss some general properties of the algebra and 
construct four different realizations of the dynamical symmetry algebra of the 
Calogero model. Using the fact that the minimal algebra of observables is 
common to the Calogero model and
the finite Chern-Simons (CS)
matrix model, we extend our analysis to the CS matrix model. We point out the 
algebraic similarities and distinctions of these models.
\end{abstract}
\vspace{1cm}
 
\maketitle

%\newpage
\section{Introduction}

The Calogero model \cite{c} is a completely integrable 
model \cite{PER} that describes a system of $N$  interacting particles in one 
dimension. Even though the interaction is highly non-trivial, one can 
construct the 
needed constants of motion, find a spectrum and  expressions
for the wave functions.
Surprisingly, 
the model and its various generalizations have been found relevant 
to a host of problems in physics (and mathematics).
It was long realized that for three special values of the interaction 
parameter $\n=1/2,1,2$ 
the model was closely related to the random matrix theory \cite{meh} of the 
three Wigner-Dyson ensembles: orthogonal, unitary and symplectic, respectively.
The particles subject to the Calogero dynamics obey fractional 
statistics \cite{ap}, 
and this  motivated investigations of the connection between
the Chern-Simons based anyonic physics in the 
fractional quantum Hall effect and the Calogero model \cite{cmat}.
The collective-field theory approach proved useful in constructing solitonic 
solutions \cite{sol}, and in establishing relations to the $d=1$ string 
theory \cite{jev}.
Recently, the interest in the Calogero model has been renewed.
It was proposed \cite{bh} 
that the  
supersymmetric extension of the  model could provide a microscopic description 
of the extremal Reissner-Nordstr\"om black hole.
The supersymmetric extensions of the Calogero model themselves 
were  analyzed in detail \cite{susy}.
Many more applications of the model have been found, which have
intensified the research of the model intrinsic properties.

Investigations of
the algebraic properties of the Calogero model in terms of
the $S_N$-extended Heisenberg algebra \cite{all} defined a basic
algebraic set-up for further research.
Floreanini et. al. \cite{2body}  showed that the dynamical symmetry algebra
of the two-body Calogero
model was a polynomial generalization of the $SU(2)$ algebra.
The  three-body problem was also treated \cite{3body},
and the  dynamical algebra of the polynomial type 
and the action of its generators on the
orthonormal basis were obtained.
It was  shown that in the  two-body case the polynomial $SU(2)$
algebra could be linearized, but an attempt  to generalize this result to the
$N$-body case  led to $(N-1)$ linear $SU(2)$ subalgebras that operated only on
subsets of the degenerate eigenspace \cite{ind}.
The general construction of the dynamical symmetry algebra was given in
Ref.\cite{before}.
Also, the bosonic realization of a non-linear symmetry algebra 
describing the structure of degenerate energy levels of the Calogero model was
obtained \cite{b2}.
In this paper we give a more detailed analysis of the algebra of observables of 
the Calogero model, discuss the spin representation of the algebra and 
present a new realization of the dynamical symmetry algebra.

The construction of the 
algebra of symmetric one-particle operators for the Calogero
model \cite{isakov} resulted in an infinite-dimensional Lie algebra, 
independent of the particle 
number and the constant of interaction. This algebra was interpreted as the 
algebra of observables for a system of identical particles  on the line. 
The  problem with fixed 
number of particles $N$, which we  discuss  here,
can be viewed as an irreducible representation of the 
afore-mentioned 
algebra. 
In Ref.\cite{poly-mx} it was claimed that this infinite-dimensional Lie algebra
was common to the matrix model. 
We argue that although this is true, the actual
models are different, as can be seen from the 
the  algebraic structure obtained for a fixed number of 
particles $N$. 

The renewed interest in the matrix model came from the 
connection with the noncommutative field theory. Namely,
following the 
Susskind conjecture \cite{suss} that the non-commutative Chern-Simons (CS)
theory provides an effective description of a quantum Hall (QH) effect,
Polychronakos
proposed \cite{pol11} 
a finite matrix version of the model. He claimed
that this 
matrix model was in fact equivalent to the Calogero model.
We analyzed 
the  algebra of observables acting on the physical Fock space of that 
model \cite{jhep}, and  observed that the minimal algebra of observables was 
identical with that of the Calogero model, 
although the complete algebraic stuctures
are  different. We identified the states in the 
$l-$th tower of the CS matrix model Fock space with the states in the physical 
Fock space of the Calogero model with the interaction parameter $\nu=l+1$.
Also,  we described quasiparticle and quasihole states in the both models
in terms of Schur functions.
Using a coherent state representation, the  wavefunctions for the
Chern-Simons matrix model proposed by Polychronakos were constructed
and compared with the Laughlin ones \cite{ks1}. The same authors also studied 
the spectrum of the model and identified the orthogonal set of states \cite{ks}.

Taking into consideration all relations between the Calogero model,
the matrix model and the QH physics,  we feel that  a more detail algebraic 
analysis is in order. 
In this paper we analyze the complete algebra of observables  of the 
CS matrix model and confirm by explicit construction that the models in 
question have different algebraic structures. 

The paper is organized as follows. In Section II we review basic steps in 
the construction of the $S_N$-extended Heisenberg algebra of the 
Calogero model 
represented on the $S_N$-symmetric subspace of the positive definite 
Fock space. In Section III we construct the minimal algebra of observables,
 ${\cal A}_N$ and discuss  its 
general properties. We establish a
mapping to the Heisenberg algebra and this gives us  a natural, orthogonal 
basis for the model.
In the followingsection
we consider the larger algebra ${\cal B}_N$ 
and show that its spin representation 
gives an irreducible representation of the 
algebra constructed in Ref.\cite{isakov}.
In  Section V we present  four 
different realizations
of the dynamical symmetry algebra 
describing the structure of the degenerate eigenspace
in detail.
The analysis of the algebra of obervables is extended to the finite 
CS matrix model in the Section VI, and in the last section we 
compare the algebraic structure of the two models. Finally, in the appendixes 
we  confirm our findings by explicit calculations.
 
\section{The Calogero model on the symmetric Fock space}

The \h \ of the (rational) Calogero model describes $N$ identical 
particles (bosons) 
interacting 
through an inverse square interaction subject to a common confining
harmonic force: 
\beq 1
H=-\frac{\hbar^2}{2m}\sum_{i=1}^N\frac{\partial^2}{\partial x_i^2}
+\frac{m\omega^2}{2}\sum_{i=1}^N x_i^2
+\frac{\n(\n-1)\hbar^2}{2m}
\sum_{i\neq j}^N\frac{1}{(x_i-x_j)^2}.\eeq
In the following we  set $\hbar$, 
the  mass of particles $m$ and the frequency of harmonic 
oscillators $\omega$ equal to one. 
The dimensionless constant $\n$ is 
the coupling constant\footnote{It is also called the statistical
parameter. Note that in one space dimension the concept of statistics requires 
a special treatment, see Refs.\cite{stat}.}
and $N$ is the number of particles.
The ground-state wave function is, up to normalization,
\beq 2
\psi_0(x_1,\ldots,x_N)=\theta(x_1,\ldots,x_N)\exp\left(-\hlf
\sum_{i=1}^N x_i^2\right),\eeq
where
\beq 3
\theta(x_1,\ldots,x_N)=\prod_{i<j}^N|x_i-x_j|^{\n},\eeq
with the ground-state energy $E_0=N[1+(N-1)\n]/2$.

The Dunkl operators \cite{D} 
$$ D_i=\partial_i+\n\sum_{j,j\neq i}^N\frac{1}{x_i-x_j}(1-K_{ij})$$
are the basic building blocks of the Calogero model creation and 
annihilation operators \cite{all}.
The elementary generators $K_{ij}$ of the symmetry group $S_N$ exchange the
labels $i$ and $j$:
\ba 5
&&K_{ij}x_j=x_iK_{ij},\; K_{ij}=K_{ji},\; (K_{ij})^2=1,\nn
&& K_{ij}K_{jl}=K_{jl}K_{il}=K_{il}K_{ij},\;{\rm for}\;i\neq j,\;
i\neq l,\;j\neq l,\nonumber\ea
and we choose $K_{ij}|0\rangle=|0\rangle$.
Now we can introduce the  creation and annihilation 
operators:
\beq 4
a_i^{\dagger}=\frac{1}{\sqrt{2}}(-D_i+x_i),\;
a_i=\frac{1}{\sqrt{2}}(D_i+x_i), \eeq
where 
the operator $a_i$ annihilates the vacuum.
Using the well-known properties of the Dunkl operators
\baa &&[D_i,D_j]=0,\;K_{ij}D_j=D_iK_{ij}, \nn &&[D_i,x_j]=
\delta_{ij}\left(1+\n\sum_{k=1}^NK_{ik}\right)-\n K_{ij},\nonumber\ea
one can easily check that the commutators of the creation and 
annihilation operators (\ref{4}) are 
\ba 6 
&&[a_i,a_j]=[a_i^{\dagger},a_j^{\dagger}]=0,\nn
&&[a_i,a_j^{\dagger}]=\left(1+\n\sum_{k=1}^NK_{ik}\right)\delta_{ij}-\n K_{ij}.
\ea
After performing a similarity transformation on the \h \ (\ref{1}), we 
obtain the reduced \h \
\beq 7
H'=\theta^{-1}H\theta=\hlf\sum_{i=1}^N\{a_i,a_i^{\dagger}\}=
\sum_{i=1}^Na_i^{\dagger}a_i+E_0 ,\eeq
acting on the space of symmetric functions. 
We restrict the Fock space $\{a_1^{\dagger n_1}\cdots a_N^{\dagger n_N}
|0\rangle\}$ 
to the $S_N$-symmetric subspace 
$F_{\rm symm}$, where ${\cal N}=\sum_{i=1}^Na_i^{\dagger}a_i$ acts as the 
total number operator. 
In the following we demand that all states should have positive norm, i. e.
$\n>-1/N$ \cite{mms}.
Next, we
 introduce the collective $S_N$-symmetric operators
\beq 8
B_n=\sum_{i=1}^Na_i^n, \; n=0,1,\ldots,N,\eeq
where $B_0$ is the constant $N$ multiplied by the identity operator, 
and $B_1$ represents 
the center-of-mass operator.
The complete $F_{\rm symm}$ can be described as $\{B_1^{\dagger n_1}
B_2^{\dagger n_2}\cdots B_N^{\dagger n_N}|0\rangle\}$. 
We wish to construct the
operators $X_k^{\dagger}$ such that $[B_1,X_k^{\dagger}]=0$
for every $k$ greater
than one, in order to separate the center-of mass coordinate.
The general solution of this equation is described by any
symmetric monomial polynomial $m_{\l}(\bar a_1,\ldots,\bar a_N)=
\sum \bar a_{1}^{\l_1}\bar a_{2}^{\l_2}\cdots \bar a_{N}^{\l_N},$
 where ${\br a}_i=a_i-
B_1/N,$
and the sum goes over all distinct permutations of $(\l_1,\l_2,\ldots,\l_N)$.
The multiset $(\l_1,\l_2,\ldots,\l_N)$ denotes any partition of ${\cal N}$
such that
$\sum_{i=1}^N\l_i={\cal N}$ and $\l_1\geq\l_2\geq\cdots\geq\l_N\geq 0$.
The "shifted" operators $\bar a_i$ and $\bar a_i^{\dagger}$ satisfy the 
following commutation relations:
\baa\label{bar6}
&&[\bar a_i,\bar a_j]=[\bar a_i^{\dagger},\bar a_j^{\dagger}]=0,\nn
&&[\bar a_i,\bar a_j^{\dagger}]=\left(1+\n\sum_{k=1}^NK_{ik}\right)\delta_{ij}
-\n K_{ij}-\frac{1}{N}.
\ea
The simplest choice of the $(N-1)$ operators commuting with the 
center-of-mass operator is
\beq 9
A_n=\sum_{i=1}^N\bar a_i^n=\sum_{i=1}^N\left(a_i-\frac{B_1}{N}\right)^n, \;
n=2,\ldots,N .\eeq
Another possible choice is $\br h_n=\sum_{\rm dist.}\bar a_{i_1}\cdots 
\bar a_{i_n}$ 
describing one-quasihole states \cite{pol1}, spanning $\{\br h_2^{\dagger}\cdots 
\br h_N^{\dagger}\}|0\rangle$.
The  symmetric Fock space $F_{\rm symm}$ is now $\{B_1^{\dagger n_1}
A_2^{\dagger n_2}\cdots A_N^{\dagger n_N}|0\rangle\}$ and after removing
the center-of-mass operator it is $\{A_2^{\dagger n_2}
\cdots A_N^{\dagger n_N}\}|0\rangle$. We have reduced the problem to the
$(N-1)$ Jacobi-type operators\footnote{Note that the algebraic sum of all
coefficients of homogeneous monomials in any $A_k$ is zero.}.
The norm of the state $\langle 0|A_n A_n^{\dagger}|0\rangle$ can be 
calculated recursively.
 We present results up to $n=4$:
\baa\label{norma}
\langle 0|A_2 A_2^{\dagger}|0\rangle &=&2(N-1)(1+N\n),\nn
\langle 0|A_3 A_3^{\dagger}|0\rangle &=&3\frac{(N-1)(N-2)}{N}(1+N\n)(2+N\n),\nn
\langle 0|A_4 A_4^{\dagger}|0\rangle &=&4\frac{(N-1)}{N^2}(1+N\n)\times\nn
&\times &\left[
6(N^2-3N+3)+N\n(5N^2-18N+18)+N^2\n^2(N-2)(N-3)\right].\ea
We see from (\ref{norma}) that for the positive definite 
Fock space  $\n$ is larger than $-1/N$.
Two different
states $A_2^{\dagger n_2}\cdots A_N^{\dagger n_N}|0\rangle$
and $A_2^{\dagger n_2'}\cdots A_N^{\dagger n_N'}|0\rangle$ with the same
energy ($\sum in_i=\sum in_i'$) are
not orthogonal. 
The total number operator on $F_{\rm symm}$ splits into
\ba a
&&{\cal N}^{\dagger}={\cal N}={\cal N}_1+{\cal\br N},\nn
&&{\cal N}_1^{\dagger}={\cal N}_1=\frac{1}{N}B_1^{\dagger}B_1, \nn
&&{\cal \br N}^{\dagger}={\cal \br N}
\equiv\sum_{k=2}^Nk{\cal N}_k .\ea
Note that ${\cal N}_k$ are the number operators of $A_k^{\dagger}$ but not of
$A_k$. Namely, $[{\cal N}_k,A_l^{\dagger}]=\delta_{kl}A_k^{\dagger}$, and
${\cal N}_k(\cdots A_k^{\dagger n_k}\cdots|0\rangle) =n_k
(\cdots A_k^{\dagger n_k}\cdots|0\rangle)$ for every $k$ larger than one, but
${\cal N}_k^{\dagger}\neq {\cal N}_k$. If ${\cal N}_k$ were hermitian, then
the eigenstates $A_2^{\dagger n_2}\cdots A_N^{\dagger n_N}|0\rangle$ would be
orthogonal, and vice versa.

\section{The ${\cal A}_N$ algebra  and bosonic realization}

Next, we discuss the  ${\cal A}_{N}$ algebra of the collective 
$S_N$-invariant
operators $A_k$ defined in Eq.(\ref{9}) and acting on $F_{\rm symm}$.
It is convenient to add  two additional operators,
$A_0=N\cdot{\bf 1}\hspace{-0.12cm}{\rm I}$ and $A_1=0$.
We  easily see that
\baa\label{Jacobi}
&&[A_i,A_j]=[A_i^{\dagger},A_j^{\dagger}]=0,\nn
&&[A_i,[A_j,X^{\dagger}]]=[A_j,[A_i,X^{\dagger}]],
\;\forall i,j,\;{\rm any\; X^{\dagger}}.\ea
The second relation in (\ref{Jacobi}) is a consequence of the Jacobi identity.
The commutator
\ba b
[A_m,A_n^{\dagger}]&=&mn\left(\sum_{i=1}^N\br a^{\dagger (n-1)}_i
\br a_i^{(m-1)}-
\frac{1}{N}A^{\dagger}_{n-1} A_{m-1}\right)\nn
&+&\sum_{k=2}^{{\rm min}(m,n)}c_k(m,n)
\left(\prod \br a^{\dagger}\right)^{n-k}
\left(\prod \br a\right)^{m-k} \ea
is an $S_N$-symmetric and  normally ordered operator.
We have separated the $c_1$ coefficient because it determines the structure 
of the algebra  of observables.
The coefficients  $c_k(m,n)$ depend also on the precise index structure (as can 
be seen in the $k=1$ case), and  the symbolical expression 
$(\prod{\cal O})^k$ denotes a product of
operators ${\cal O}_i$  of the total order $k$ in $\br a_i(\br a_i^{\dagger})$.
Hence, the structure of the ${\cal A}_{N}(\n)$ algebra is of the
following type:
\beq d
[A_{i_1},[A_{i_2},\ldots,[A_{i_j},A_j^{\dagger}]]\cdots]=\sum(\prod A)^{I-j},
\eeq
where $I=\sum_{\alpha=1}^ji_{\alpha}\geq 2j$, and similarly for the hermitian 
conjugate case.
Generally, $j$ successive commutators of $A_{i_1},\ldots,A_{i_j}$ with
$A_j^{\dagger}$, form  a homogeneous polynomial $\sum(\prod A)^{I-j}$ in
$a_i$ of order $I-j$ with coefficients independent  of $\n$.
Therefore, we stress that the algebraic relations (\ref{d})
are
common to all sets of operators $\{A_k,A_k^{\dagger}\}$, with $k=2,3,\ldots,N$,
satisfying
\bee\label{cmn}
[A_i,A_j]=[A_i^{\dagger},A_j^{\dagger}]=0,\; [{\cal N}_i,A_j^{\dagger}]=
\delta_{ij}A_j^{\dagger}.\eeq
Two different sets of operators satisfying the same
algebra (\ref{d}) differ only in the generalized vacuum conditions, see below.
So, we denote the common algebra of the operators
$\{A_k,A_k^{\dagger}\}$  by ${\cal A}_N$, and its representation
for a given $\n>-1/N$ by ${\cal A}_N(\n)$.

The term $A_{I-j}$ on the r.h.s. of Eq.(\ref{d}) appears with the 
coefficient\footnote{ Note that the choice of one-quasihole states for the
basis of the symmetric Fock space would
result in a similar algebraic structure, but with different coefficients
on the r.h.s. of Eq.(\ref{d}).}
$(\prod_{\alpha}i_{\alpha})j!$.  
There are ${2N-1\choose N} -N$ linearly independent relations (\ref{d}).
Specially, we find
\bee\label{Jss}
[\underbrace{A_2,[A_2,\ldots[A_2}_n,A_n^{\dagger}]\ldots]=2^nn!A_n\;,\eeq
and more generally
\bee\label{Jss2}
[\underbrace{A_2,[A_2,\ldots[A_2}_m,[\underbrace{A_2^{\dagger},[A_2^{\dagger}
,\ldots[A_2^{\dagger}}_m,A_n]\ldots]=(-)^m2^{2m}\frac{m!n!}{(n-m)!}A_n.\eeq
Relations ({\ref{Jss}) and ({\ref{Jss2}) can be proved 
 using induction and Jacobi identities.
Note that any $A_n,\;n>N$, can be algebraically  expressed in terms of
$A_m,\;m\leq N$, see Appendix A.
Hence, the ${\cal A}_N$ algebra expressed in terms of $2(N-1)$
algebraically independent operators, is closed, finite and of the 
polynomial type.

For general $N$, the ${\cal A}_N(\n)$  algebra  of the 
collective $S_N$-symmetric
operators (\ref{9})
completely determines the action of $A_k$  on any state: 
\bee\label{nesto}
A_kA_2^{\dagger n_2}\cdots A_N^{\dagger n_N}|0\rangle=
\sum \left(\prod A^{\dagger}\right)^{{\cal N}-k}|0\rangle,\;k\leq {\cal N}.\eeq
In order to calculate the precise form of the r.h.s. of Eq.(\ref{nesto}), we
apply the hermitian conjugate relation Eq.(\ref{d}) on the l.h.s. of
Eq.(\ref{nesto}) shifting the operator $A_k$ to the right, at least by  one
place. We repeat this iteratively as long as the number of $A^{\dagger}$'s
on the right from $A_k$
is larger or equal to the index $k$.
For
$k>\sum n_i$, we calculate the finite set of relations,
so-called generalized vacuum conditions, directly 
from (\ref{9}) and
(\ref{bar6}).
We show that  the {\it minimal} set of generalized vacuum conditions
needed to completely
define the representation of the algebra (\ref{d}) on the Fock space is
\baa\label{cV}
A_2A_2^{\dagger}\vac &=&2(N-1)(1+\n N)\vac,\nn
A_3A_3^{\dagger}\vac &=&3(N-1)(N-2)(1+\n N)(2+\n N)/N\vac,\nn
A_3A_3^{\dagger 2}\vac &=&3(N-2)(2+\n N)[2(N-1)(1+\n N)+18]/N\vac.\ea
Namely,
the operators $A_2,\;A_3$ and hermitian conjugates play a distinguished role in
the algebra, as all other operators $A_n,\;A_n^{\dagger}$ for $n\geq 4$ can
be expressed as successive commutators (\ref{d}), using only $A_2,\;A_3$ and
their hermitian conjugates. Therefore, one can  derive all other
generalized vacuum conditions using (\ref{d}) and (\ref{cV}).

Note that the action
of $A_iA_j^{\dagger}$ (and ${\cal N}_i$)
 on the symmetric Fock subspace can be written as an
infinite, normally ordered expansion
\bee\label{mmm}
A_iA_j^{\dagger}=\sum_{k=0}^{\infty}\left(\prod A^{\dagger}\right)^{k+j}
\left(\prod A\right)^{k+i},\; \forall i,j.\eeq
Applied to a monomial state of the finite order  ${\cal N}$
 in $F_{\rm symm}$, only the finite number of terms
in Eq.(\ref{mmm}) will contribute.
For the $N=3$ case, we provide explicit calculations in Appendix B, 
demonstrating 
the main features of the ${\cal A}_N(\n)$ algebra.

There is a useful (and orthogonal) basis for the  problem at hand.
We define the operators $\A_i,\;i=2,\ldots,N$, satisfying the
vacuum condition $\A_i|0\rangle=0$ and
\beq s
\A_iA_2^{\dagger n_2}\cdots A_i^{\dagger n_i}\cdots A_N^{\dagger n_N}|0\rangle=
n_iA_2^{\dagger n_2}\cdots A_i^{\dagger (n_i-1)}\cdots
A_N^{\dagger n_N}|0\rangle
,\eeq
for all $n_2,\ldots,n_N\in \Nb_0$.
As a consequence of $[A_i^{\dagger},A_j^{\dagger}]=0$, we immediately find
\beq z
[\A_i,\A_j]=0,\;[\A_i,A_j^{\dagger}]=\delta_{ij},\;i,j=2,\ldots ,N.\eeq
We define a dual Fock space as a set of states
$\langle 0|\A_2^{n_2}\cdots \A_N^{n_N},\;n_i\in \Nb_0$, such that
\bee\label{dualF}
\langle 0|\A_2^{n'_2}\cdots \A_N^{n'_N}A_2^{\dagger n_2}\cdots
A_N^{\dagger n_N}|0\rangle=\prod_in_i!\delta_{n_in'_i}.\eeq
Although the operators $\A_i$ and $A_i^{\dagger}$ are not hermitian
conjugate, the second equation in (\ref{z}) induces a new scalar product
with respect to
which the states $A_2^{\dagger n_2}\cdots A_N^{\dagger n_N}|0\rangle$ are
orthogonal to $\langle 0|\A_2^{n_2}\cdots \A_N^{n_N}$, according to
Eq.(\ref{s}). Generally, the operators $\A_i$ can be written in the form
\bee\label{gen3}
\A_i=A_i+\sum_{k=2}^{\infty}(\prod A^{\dagger})^k(\prod A)^{k+i}.\eeq
We give an example for the $N=3$ case in Appendix B.
Then we define the number operators ${\cal N}_i$ and the transition
number operators ${\cal N}_{i,j}$ as
\ba x
&&{\cal N}_{i,j}
=A_i^{\dagger}\A_j,\; {\cal N}_i={\cal N}_{i,i}=A_i^{\dagger}\A_i,\;
{\cal N}_1=\frac{1}{N}B_1^{\dagger}B_1,\nn
&&[{\cal N}_{i,j},A_k^{\dagger}]=\delta_{jk}A_i^{\dagger},\;
[{\cal N}_{i,j},\A_k]=-\delta_{ik}\A_j,\;
[{\cal N}_{i,j},{\cal N}_{k,l}]=\delta_{jk}{\cal N}_{i,l}-\delta_{il}
{\cal N}_{k,j}.\ea
One can define a new hermitian conjugation operation
 $(^*)$ in the following way:
$${\cal N}_{i,j}^*={\cal N}_{j,i},\;(A_k^{\dagger})^*=\A_k,\;
(\A_k)^*=A_k^{\dagger},\;B_1^*=B_1^{\dagger},\;i,j,k=2,\ldots,N.$$
This realization of the algebra of observables provides an orthogonal 
basis in the dual Fock space and 
leads to a simple realization of the dynamical symmetry algebra.

We point out that if
 the set of operators $\{A_k,A_k^{\dagger}\}$ satisfies relations
(\ref{cmn}), then there exists the  mapping $A_i=f(b_i,b_i^{\dagger})$
from the ordinary Bose oscillators
$\{b_i,b_i^{\dagger}\}$ to $\{A_k,A_k^{\dagger}\}$.
If there are no null-norm vectors in the $\{A\}$ Fock space, there exists the
inverse mapping $b_i=f^{-1}(A_k,A_k^{\dagger})$. It has been found in 
Ref.\cite{mms} that the full Fock space $\{a_1^{\dagger n_1},\ldots,
a_N^{\dagger n_N}|0\rangle\}$ for $\n >-1/N$ does not contain any additional 
null-norm states when compared with the $\{b\}$ Fock space. 
As $F_{\rm symm}$ 
is a subspace of the full Fock space, it also does not contain any additional 
null-norm states, so $F_{\rm symm}$ is isomorphic to the $\{b\}$ Fock space, 
and the mapping 
$f$ is invertible for  $\n >-1/N$.
The point $\nu=-1/N$ is a critical point of the algebra and a description 
of the system near  this point is given in Ref.\cite{krit}.

In our case of interest, namely in the $A_N(\n)$ algebra, we have started with
 positive-norm states, i.e.,
$\n >-1/N$, so there exists a real mapping $f$ and its inverse $f^{-1}$.
In general, one can write the infinite series as
\baa\label{gener}
A_n&=&\sum\left(\prod b_i^{\dagger n_i}\right)^k\left(\prod b_j^{n_j}\right)
^{k+n},\;\sum in_i=k,\;\sum jn_j=k+n,\nn
b_n&=&\sum\left(\prod A^{\dagger}\right)^k\left(\prod A\right)^{k+n},\ea
and then calculate the coefficients.
There is     freedom in mapping (\ref{gener}), which appears
in the subspaces spanned by the monomials $(\prod A^{\dagger})
^{\cal\br N}$ (or $(\prod b^{\dagger})^{\cal\br N})$
of the same order ${\cal\br  N}$.
A simple and natural choice of fixing  these  boundary conditions is
\baa\label{gfix}
&&A_2^{\dagger n_2}|0\rangle =\sqrt{\frac{\langle 0|A_2^{n_2}
A_2^{\dagger n_2}|0\rangle}{n_2!}}b_2^{\dagger n_2}|0\rangle ,\nn
&&A_3^{\dagger n_3}|0\rangle \sim \sum\left(\prod b_2^{\dagger}b_3^{\dagger}
\right)^{n_3}|0\rangle ,\nn
&&\vdots \nn
&&A_N^{\dagger n_N}|0\rangle \sim \sum\left(\prod b_2^{\dagger}\cdots
b_N^{\dagger}
\right)^{n_N}|0\rangle ,\ea
and generally,
\bee\label{gfixg}
A_2^{\dagger n_2}A_3^{\dagger n_3}\cdots A_N^{\dagger n_N}|0\rangle \sim \sum
b_2^{\dagger n_2}\sum\left(\prod b_2^{\dagger}b_3^{\dagger}\right)^{n_3}
\cdots\left(\prod b_2^{\dagger}\cdots b_N^{\dagger}\right)^{n_N}|0\rangle .\eeq
Only after fixing this freedom, one can determine the coefficients
 in Eq.(\ref{gener}) in a unique way.
For the N=3 case, we present results for the first few coefficients in
Eq.(\ref{gener}),  up to $k+n\leq 5$, for the
operators $A_2$ and $A_3$ in Appendix B.
The states in the
$\{A\}$ Fock space are not orthogonal. However, the monomial states
$\prod b_i^{\dagger n_i}/\sqrt{n_i!}|0\rangle$
in the $\{b\}$ Fock space are orthogonal, so when we express
$b_i=f^{-1}(A_k,A_k^{\dagger})$, we obtain  natural orthogonal states in
the $\{A\}$ Fock space, labelled by $(n_2,\ldots,n_N)$, i.e. by free
oscillator quantum numbers.
Degenerate, orthogonal energy eigenstates of
level ${\cal N}$ are then defined by ${\cal N}=\sum in_i$.

\section{The ${\cal B}_N(\n)$ algebra and spin representation}

One can construct the larger closed ${\cal B}_N(\n)$ algebra  containing 
${\cal A}_N(\n)$ as a subalgebra. This larger algebra appears naturally when
one calculates the commutators between the operators $A_i$ and $A_j^{\dagger}$, 
defined in Eq.(\ref{9}). 
We discussed this algebra in Ref.\cite{before}; here we  repeat the 
main results for completeness and present an alternative construction of the 
algebra. 

One can define $S_N$-symmetric operators
$\br B_{n,m}$:
\beq g 
\br B_{n,m}=\sum_{i=1}^N\bar a_i^{\dagger n}\bar a_i^m=\br B^{\dagger}_{m,n},\;
n,m\in \Nb_0.\eeq
The operators $\br B_{n,m}$ can be represented in the symmetric Fock space:
$$\br B_{n,m}A_2^{\dagger n_2}\cdots A_N^{\dagger n_N}|0\rangle\equiv
\br B_{n,m}(\prod A^{\dagger})^{\cal N}|0\rangle=\sum (\prod A^{\dagger})^
{{\cal N}+n-m}|0\rangle.$$

There are $1/2(N+4)(N-1)$ algebraically independent $\br B_{n,m}$ operators 
contained in the algebra ${\cal B}_N(\n)$,
namely
$2(N-1)$ operators
$\br B_{n,0}=A_n^{\dagger},\;{\rm for}\;n=2,3,\ldots,N$
and their hermitian conjugates, and
$N(N-1)/2$ operators $\br B_{n,m},\;{\rm for}\;
n,m\geq 1, n+m\leq N$. 
One can express the operators  $\br B_{n,m}$
for $n+m>N$ in terms of the algebraically independent operators $\br B_{n,m}$
with $n+m\leq N$, see Appendix A.
This is a consequence of Cayley-Hamilton theorem.
Generally,
\ba h
&&\br B_{1,1}={\cal\br N},\;
[B_1, \br B_{n,m}]=0,\;[\br B_{1,1}, \br B_{n,m}]=(n-m)\br B_{n,m},\nn
&&[\bar a_i,\br B_{n,m}]=n\left[\bar a_i^{\dagger (n-1)}\bar a_i^m-\frac{1}{N}
\br B_{n-1,m}\right].\ea

In the case of $N$ free harmonic osillators with $\n=0$, 
we find the general ${\cal B}_N(0)$-algebra relation
\ba j
&&[\br B_{m',m},\br B_{n,n'}]=\sum_{k=1}^{{\rm min}(n,m)}\beta_k(m,n)
\left(\frac{1}{N}\right)^k
\left\{\left[(N-1)^k+1\right]
\br B_{m'+n-k,m+n'-k}\right. \nn 
&&-\left. \br B_{m',m-k}\br B_{n-k,n'}+
\sum_{s=1}^{{\rm min}(n-k,m-k)}\beta_s(m-k,n-k)\br B_{m'+n-k-s,m+n'-k-s}
\left(1-\frac{1}{N}\right)^s \right\}\nn
&&-\{m'\leftrightarrow n,n'\leftrightarrow m\},
\ea
where $$\beta_k(m,n)=\frac{m!n!}{k!(m-k)!(n-k)!}.$$

For $\n\neq 0$, the structure of the ${\cal B}_N(\n)$ algebra becomes more 
complicated. 
New polynomial terms of the form $(\prod_{\alpha}\br B_{n_{\alpha},m_{\alpha}})$
with $\sum_{\alpha}n_{\alpha}\leq n+m'-1,\;\sum_{\alpha}m_{\alpha}
\leq n'+m-1$ appear on the r.h.s. of the commutation relation (\ref{j}). 
The corresponding coefficients are polynomial in $\n$, 
vanishing when $\n$ goes to zero. The coefficients of the leading terms 
($k=1$ in Eq.(\ref{j}))
do not depend on $\n$, i. e. they are the same  for any $\n $.
For example, for arbitrary $N$ and $\n$, we find
\beq l
\left[A_2,A_n^{\dagger}\right]=2n\br B_{n-1,1}
+n\left(\frac{N-1}{N}\right)A_{n-2}^{\dagger}(n-1+\n N)
+n\n\sum_{i=1}^{n-2}\left(A_{n-2-i}^{\dagger}A_i^{\dagger}
-A_{n-2}^{\dagger}\right),\eeq
It is known that
for $N\rightarrow\infty$ and $\n=0$ the corresponding algebra is
$W_{1+\infty}$, so it would be interesting to see what kind of
deformation the ${\cal B}_{\infty}(\n)$
algebra (for nontrivial $\n$) represents.
Some investigation in this direction has already been done, see Refs.\cite{w}.

Alternatively, the ${\cal B}_N(\n)$ algebra can be constructed by
grouping the generators into  $su(1,1)$-spin multiplets. Note that the operators
\bee\label{sl2}
J_+=\hlf A_2^{\dagger},\; J_-=\hlf A_2,\; J_0=\frac{1}{8}[A_2,A_2^{\dagger}]
\eeq
 generate
the $sl(2)$-algebra. The complete set of generators
spanning the ${\cal B}_N(\n)$ algebra is given by $(N-1)$  non-degenerate
spin multiplets with $s=1,3/2,2,\ldots ,N/2$. The unique generator with spin
 $s$  and projection $s_z$ is defined as
\bee\label{sp1}
 J_{s,s_z}=\frac{1}{2^{(s-s_z)}}\sqrt{\frac{(s+s_z)!}
{(s-s_z)!}}
 [\underbrace{A_2,
\cdots,[A_2}_{(s-s_z)},A_{2s}^{\dagger}]\cdots].\eeq
Its hermitian conjugate is simply $J_{s,s_z}^{\dagger}=J_{s,-s_z}$.
One can show that by proving relation (\ref{Jss2})
$$[\underbrace{A_2,[A_2,\ldots[A_2}_{s-s_z},
[\underbrace{A_2^{\dagger},[A_2^{\dagger}
,\ldots[A_2^{\dagger}}_{s-s_z},A_{2s}]\ldots]=(-)^{(s-s_z)}2^{2(s-s_z)}
\frac{(s-s_z)!(2s)!}{(s+s_z)!}A_{2s},$$
by induction and using Jacobi identities.
The  action of  elements of the $sl(2)$-algebra on the generator $ J_{s,s_z}$
is defined in the
following way:
\baa\label{sp3}
\left[J_-,J_{s,s_z}\right]&=&\sqrt{(s+s_z)(s-s_z+1)}J_{s,s_z-1},\nn
\left[J_+,J_{s,s_z}\right]&=&-\sqrt{(s-s_z)(s+s_z+1)}J_{s,s_z+1},\nn
\left[J_0,J_{s,s_z}\right]&=&s_zJ_{s,s_z}. \ea
Let us write down the first few spin multiplets for any $N$ in terms of 
$\br B_{i,j}$ operators:

$s=1,\;N\geq 2$:
\baa\label{sip1}
J_{1,1}&=&\sqrt{2}A_2^{\dagger}=2\sqrt{2}J_+=J_{1,-1}^{\dagger},\nn
J_{1,0}&=&2\br B_{1,1}+(N-1)(1+N\n)=4J_0,\ea

$s=\frac{3}{2},\; N\geq 3$:
\bee\label{sp32}
J_{\frac{3}{2},\frac{3}{2}}=\sqrt{6}A_3^{\dagger}=
J_{\frac{3}{2},-\frac{3}{2}}^{\dagger},\;
J_{\frac{3}{2},\frac{1}{2}}=3\sqrt{2}\;\br B_{2,1}=
J_{\frac{3}{2},-\frac{1}{2}}^{\dagger},\;
\eeq

$s=2,\; N\geq 4$:
\baa\label{sp2}
J_{2,2}&=&2\sqrt{6}A_4^{\dagger}=J_{2,-2}^{\dagger},\nn
J_{2,1}&=&2\sqrt{6}\left\{2\br B_{3,1}
+\left[3\left(\frac{N-1}{N}\right)+\n(2N-3)
\right]A_2^{\dagger}\right\}=J_{2,-1}^{\dagger},\nn
J_{2,0}&=&\sqrt{2}\left\{6\br B_{2,2}+\br B_{1,1}\left[12
\left(\frac{N-1}{N}\right)
+\n(7N-12)\right]\right.\nn 
&+&\left.(N-1)(1+N\n)\left[3\left(\frac{N-1}{N}\right)+\n(2N-3)\right]
\right\}.\ea

The algebra of the operators $J_{s,s_z}$ is finite, closed and of general form:
\bee\label{jg}
\left[J_{s_1,s_{1z}},J_{s_2,s_{2z}}\right]=\sum \left(\prod 
J\right)_{S,s_{1z}+s_{2z}}, \eeq
where $|s_{1z}+s_{2z}|\leq S\leq s_1+s_2-1$.
Using the
definition (\ref{sp3}) and relation (\ref{sip1}),
one can easily obtain commutation relations involving operators 
of the $s=1$ spin multiplet, 
\bee\label{sp1c}
[J_{1,0},J_{s,s_z}]=4s_zJ_{s,s_z},\;[J_{1,\pm 1},J_{s,s_z}]=\mp
\sqrt{8(s\mp s_z)(s\pm s_z+1)}J_{s,s_z\pm 1}.\eeq
For arbitrary $N$, the algebra of $s=3/2$ operators is
\baa
\left[J_{\frac{3}{2},\frac{1}{2}},J_{\frac{3}{2},\frac{3}{2}}\right]&=&
\frac{9}{\sqrt{2}}J_{2,2}-\frac{9\sqrt{3}}{N}J_{1,1}^2,\nn
\left[J_{\frac{3}{2},-\frac{1}{2}},J_{\frac{3}{2},\frac{3}{2}}\right]&=&
\frac{9}{\sqrt{2}}J_{2,1}-\frac{9\sqrt{2}}{N}J_{1,1}J_{1,0}+9\sqrt{6}J_{1,1}
\left[-\frac{2}{N}+\n(2-N)\right],\nn
\left[J_{\frac{3}{2},-\frac{3}{2}},J_{\frac{3}{2},\frac{3}{2}}\right]&=&
\frac{9}{\sqrt{2}}J_{2,0}-\frac{27}{N}J_{1,1}J_{1,-1}-9\left(\frac{6}{N}+
\frac{N}{2}\n\right)J_{1,0}+a,\nn
\left[J_{\frac{3}{2},-\frac{1}{2}},J_{\frac{3}{2},\frac{1}{2}}\right]&=&
\frac{9}{\sqrt{2}}J_{2,0}+\frac{9}{N}J_{1,1}J_{1,-1}-\frac{18}{N}J_{1,0}^2
+18J_{1,0}
\left[\frac{1}{N}+\left(\frac{3}{4}N+4\right)\right]+b,\nonumber\ea
where the constants $a$ and $b$ are
\baa
a&=&+9(N-1)(1+N\n)\left[\frac{N+1}{N}+\left(\frac{N}{2}-1\right)\n\right],\nn
b&=&-18(N-1)(1+N\n)\left[-\hlf+\frac{3}{2N}+N+\left(N^2-\frac{11}{4}N-\hlf
\right)\n\right].\nonumber \ea

For $N=3$,  the ${\cal B}_3(\n)$ algebra (including $s=1$ and $s=3/2$
spin-mutiplets) is in
full agreement with Ref.\cite{3body}. The exact
correspondence between $Y_s$ and $J$
defined in Ref.\cite{3body} and our operators
$J_{s,s_z}$ is
\baa\label{corr}
&&Y_1=-\sqrt{2}J_{1,1},\;Y_{3/2}=2J_{\frac{3}{2},\frac{3}{2}},\;
Y_{1/2}=-2\sqrt{3}J_{\frac{3}{2},\frac{1}{2}},\nn
&&J=\frac{1}{4}\left(J_{1,0}-2\right),\;
J_y^2=\frac{1}{16}\left(J_{1,0}-2\right)^2-\frac{1}{8}J_{1,1}J_{1,-1}.\ea
This  representation of the algebra of observables 
can be viewed as a generalization of the
polynomial algebras for $N=2$ (Ref.\cite{2body}) and $N=3$ (Refs.\cite{3body}
 and \cite{Sn}) to
the general $N$.

The ${\cal B}_N(\n)$ represents  an irreducible representation of 
the infinite-dimensional Lie algebra ${\cal G}$ of 
all possible strings of consecutive 
commutators $[B_{i_n},[B_{i_{n-1}},\ldots[B_{i_2},B_{i_1}]\ldots]$ introduced 
in Ref.\cite{isakov}. The ${\cal G}$ algebra  depends neither  on the particle 
number $N$  nor on the constant of interaction $\n$.
The elements of the 
algebra ${\cal G}$ fall into spin multiplets. There is a unique spin
multiplet with a maximal spin $s_{\rm max}=(m+n)/2$, no spin multiplet
with spin $s=s_{\rm max}-1$, and many spin multiplets with $s<s_{\rm max}-1$.
On the other hand, we have started with a 
fixed number $N$ of Calogero particles 
(oscillators) and an
arbitrary $\n>-1/N$. Then we have defined a 
finite closed algebra ${\cal B}_N(\n)$ of
operators $\br B_{m,n}$, with $m+n\leq N$. We have shown (see Appendix A) that
all operators $\br B_{m,n}$, with $m+n\geq N$ can be expressed in terms of
$\br B_{m,n}$ with $m+n\leq N$. These operators have a unique $su(1,1)$
decomposition into unique spin multiplets with spin $s=1,3/2,\ldots,N/2$.
Degenerate spin multiplets in the approach of Isakov and Leinaas \cite{isakov}
are just  composites of lower-spin multiplets in our picture.

\section{Dynamical symmetry of the Calogero model}

The dynamical symmetry algebra 
${\cal C}_N(\n)$ of the Calogero model is defined as 
maximal algebra commuting with the \h \ (\ref{7}), on the restricted Fock 
space $F_{\rm symm}$. The generators of the ${\cal C}_N(\n)$  algebra
act among the degenerate states with a fixed energy $E={\cal N}
+E_0,$
${\cal N}$ a non-negative integer. 
Starting from any of the degenerate states with energy $E$, 
all other states can be reached by applying the generators of the 
algebra.
Degeneracy appears for ${\cal N}\geq 2$. The vacuum $|0\rangle$ and the 
first excited state $B_1^{\dagger}|0\rangle$ are non-degenerate. 
For ${\cal N}=2$, 
the degenerate states are $B_1^{\dagger 2}|0\rangle$ and $A_2^{\dagger}
|0\rangle$; for ${\cal N}=3$, the degenerate states are 
$B_1^{\dagger 3}|0\rangle$, $B_1^{\dagger}A_2^{\dagger}|0\rangle$ and 
$A_3^{\dagger}|0\rangle$, etc. The number of degenerate states of level 
${\cal N}$ 
is given by partitions ${\cal N}_1,\ldots,{\cal N}_k$ of ${\cal N}$ such that 
${\cal N}=\sum_k k{\cal N}_k$.
The generators of the algebra
${\cal C}_N(\n)$ can be chosen in different ways, and in the following 
we  present four different sets of generators. 

{\bf Set 1}. 
Let us choose $1/2(N+4)(N-1)$ algebraically independent
 generators $X_{i,j},\;(i+j\leq N)$ in the following way:
\beq m
X_{i,j}=\br B_{i,j}\left(\frac{B_1}{\sqrt{N}}\right)^{(i-j)},\;
X_{j,i}=X_{i,j}^{\dagger}=\br B_{j,i}\left(\frac{B_1^{\dagger}}
{\sqrt{N}}\right)^{(i-j)}, \; i\geq j.\eeq
For example,
\baa
&&X_{i,0}=A_i^{\dagger}\left(\frac{B_1}{\sqrt{N}}\right)^{i},\;
X_{0,i}=A_i\left(\frac{B_1^{\dagger}}{\sqrt{N}}\right)^{i},\nn
&&[X_{i,0},X_{j,0}]=[X_{0,i},X_{0,j}]=0,\; X_{i,i}^{\dagger}=X_{i,i}
=\br B_{i,i}.\ea
The generators $X_{i,i}$ are hermitian but they do not commute because 
$\br B_{i,i}$'s do not commute even for $\n=0$ \cite{before}. 
On the other hand, the number 
operators ${\cal N}_k$ (\ref{a}) commute but are not hermitian because
 the states 
$A_2^{\dagger n_2}\cdots A_N^{\dagger n_N}|0\rangle$ are not mutually 
orthogonal.
Generally,
\ba r
&&\left[{\cal N}_1,X_{i,j}\right]=-(i-j)X_{i,j},\; \left[{\cal \br N},X_{i,j}
\right]=(i-j)X_{i,j},\nn
&& \left[H,X_{i,j}\right]=\left[{\cal N},X_{i,j}\right]=0,
\; {\rm for\; all\; i,j.}\ea

The general structure of the commutation relations for $i\geq j,\;k\geq l$ is
\beq n
\left[X_{i,j},X_{k,l}\right]=\left[\br B_{i,j},\br B_{k,l}\right]
\left(\frac{B_1}{\sqrt{N}}\right)
^{(i-j)+(k-l)}=\sum\left[\prod_{\alpha}X_{n_{\alpha},m_{\alpha}}
\right] ,\eeq
and for $i>j,\;k<l$,
\beq o
\left[X_{i,j},X_{k,l}\right]=\sum\left[\prod_{\alpha}X_{n_{\alpha},m_{\alpha}}
g_{n_{\alpha},m_{\alpha}}({\cal N}_1)\right]
+X_{i,j}X_{k,l}f_{ijkl}({\cal N}_1),\eeq
with the restriction $0\leq \sum m_{\alpha}\leq j+l-1
,\;0\leq \sum n_{\alpha}\leq i+k-1$, and similarly for hermitian conjugate 
relations. The functions $f$ and $g$ 
are generally rational functions of ${\cal N}_1$, with the finite action on all
states.
One can show that for $i\geq j$:
\baa \left(\frac{B_1}{\sqrt{N}}\right)^{i}\left(\frac{B_1^{\dagger}}
{\sqrt{N}}\right)^{j}&=&\left(\frac{B_1}{\sqrt{N}}\right)^{(i-j)}
({\cal N}_1+1)\cdots({\cal N}_1+j)\nn &=&({\cal N}_1+1+i-j)\cdots({\cal N}_1+i)
\left(\frac{B_1}
{\sqrt{N}}\right)^{(i-j)},\nonumber\ea
and
\baa \left(\frac{B_1^{\dagger}}{\sqrt{N}}\right)^{j}\left(\frac{B_1}
{\sqrt{N}}\right)^{i}&=&{\cal N}_1
({\cal N}_1-1)\cdots({\cal N}_1-j+1)\left(\frac{B_1}
{\sqrt{N}}\right)^{(i-j)}\nn &=&\left(\frac{B_1}{\sqrt{N}}\right)^{(i-j)}
({\cal N}_1-i+j)\cdots({\cal N}_1-i+1), \nonumber\ea
and similarly for $i<j$. Now it is easy to see that
\bee \left[\left(\frac{B_1}{\sqrt{N}}\right)^{i},\left(\frac{B_1^{\dagger}}
{\sqrt{N}}\right)^{j}\right]=\sum_{k=1}^{{\rm min}(i,j)}\beta_k(i,j)
\left(\frac{B_1^{\dagger}}{\sqrt{N}}\right)^{(j-k)}\left(\frac{B_1}{\sqrt{N}}
\right)^{(i-k)}.\eeq

The ${\cal C}_N(\n)$ algebra is intrinsically polynomial.  
For $N=2$, the ${\cal C}_2(\n)$-Calogero
algebra is the $SU(2)$-polynomial (cubic) algebra \cite{2body}, i.e.
$[X_{2,0},X_{0,2}]=P_3({\cal N}_1,{\cal \br N})$. In this case, the  
${\cal C}_2(\n)$  algebra 
 can be linearized to the ordinary $SU(2)$ algebra 
owing to the fact that there are two independent, uncoupled oscillators $B_1$ 
and $A_2$, which can be mapped to two ordinary Bose oscillators \cite{mmp}.
The $SU(2)$ generators are
\baa\label{su2}
{\cal J}_+&=&
\frac{1}{4\sqrt{({\cal N}_1-1)({\cal \br N}+1+2\n)}}B_1^{\dagger 2}A_2,\nn
{\cal J}_-&=&
A_2^{\dagger}B_1^2\frac{1}{4\sqrt{({\cal N}_1-1)({\cal \br N}+1+2\n)}}=\left(
{\cal J}_+\right)^{\dagger},\nn
{\cal J}_0&=&
\frac{1}{16}\left(\frac{1}{({\cal N}_1-1)}B_1^{\dagger 2}B_1^2-
\frac{4}{({\cal \br N}-1+2\n)} A_2^{\dagger}A_2\right)=\frac{1}{4}
\left({\cal N}_1-{\cal \br N}\right),
\ea
satisfying $[{\cal J}_+,{\cal J}_-]=2{\cal J}_0,\;[{\cal J}_0,{\cal  J}_{\pm}]
=\pm {\cal J}_{\pm}$.
The generators ${\cal J}_+$ and ${\cal J}_-$ 
are hermitian conjugates to each other and in 
this respect differ from the construction done in Ref.\cite{ind}.
For $N=3$, the ${\cal C}_3(\n)$ algebra in Eqs.(\ref{n},\ref{o}) 
is the same as in Ref.\cite{3body}. One can easily find the 
exact correspondence 
using Eq.(\ref{corr}). 

{\bf Set 2}. We can construct  a new set of generators of the dynamical 
${\cal C}_N(\n)$  algebra 
in terms of the operators $J_{s,s_z}$.
For general $N$, we define
\baa\label{spgen}
\tilde X_{s,s_z}&=& J_{s,s_z}
\left(\frac{B_1}{\sqrt{N}}\right)^{2s_z}, \; {\rm for}\;
s_z\geq 0,\nn
\tilde X_{s,s_z} &=&
J_{s,s_z}\left(\frac{B_1^{\dagger}}{\sqrt{N}}\right)^{-2s_z},\;
{\rm for}\; s_z<0.\ea 
The operators $\tilde X_{s,s_z}$ satisfy a similar
algebraic relation as the generators of the preceding realization,
 Eqs.(\ref{n}) and (\ref{o}): 
\bee\label{nn}
\left[\tilde X_{s,s_z},\tilde X_{s',s_z'}\right]=
\left[J_{s,s_z},J_{s',s_z'}\right]
\left(\frac{B_1}{\sqrt{N}}\right)
^{2(s_z+s_z')}=\sum\left(\prod \tilde X
\right)_{S,s_z+s_z'},\;s_z,s_z'\geq 0 ,\eeq
where $|s_z+s_z'|\leq S\leq s+s'-1$,
and for $s_z\geq 0,\;s_z'<0$:
\bee\label{oo}
\left[\tilde X_{s,s_z},\tilde X_{s',s_z'}\right]
=\sum\left[\prod_{\alpha}\tilde X_{s^{\alpha},s_z^{\alpha}}
\tilde g_{s^{\alpha},s_z^{\alpha}}({\cal N}_1)\right]
+\tilde X_{s,s_z}\tilde X_{s',s_z'}\tilde f_{s,s_z,s',s_z'}({\cal N}_1),\eeq
with the restriction $\sum s_z^{\alpha}=s_z+s_z'
,\;\sum s^{\alpha}\leq s+s'-1$, and similarly for hermitian conjugate
relations. The functions $\tilde f$ and $\tilde g$
are generally rational functions of ${\cal N}_1$, with the finite action on all
states.
For general $N$, we present several typical commutators that demonstrate the
general structure given by Eqs.(\ref{nn},\ref{oo}):
\baa\label{tilex}
\left[\tilde X_{1,0},\tilde X_{s,s_z}\right]&=&4s_z\tilde X_{s,s_z},\nn
\left[\tilde X_{1,1},\tilde X_{s,s_z}\right]&=&
-\sqrt{8(s-s_z)(s+s_z+1)}\tilde X_{s,s_z+1},\;s_z\geq 0,
\nn \left[\tilde X_{1,-1},\tilde X_{s,s_z}\right]&=&+\sqrt{8(s+s_z)(s-s_z+1)}
\tilde X_{s,s_z-1}({\cal N}_1-2s_z+2)({\cal N}_1-2s_z+1)\nn &-&\frac{2s_z}
{{\cal N}_1-2s_z}
\tilde X_{1,-1}\tilde X_{s,s_z}\left(2+\frac{2s_z-1}{{\cal N}_1-2s_z-1}\right),
\;s_z>0.\ea 
This  construction can be viewed as a generalization of the
polynomial algebras for $N=2$ \cite{2body} and $N=3$ \cite{3body} to 
the general $N$, using the ${\cal B}_N(\n)$ algebra.

{\bf Set 3}. Here we introduce a new set of generators of the dynamical 
 ${\cal C}_N(\n)$ algebra
 which obey very simple relations and possess interesting 
properties.
The dynamical algebra can be  defined through the  set
of $N^2$ generators in terms of the transition number operators (\ref{x}) 
\ba w
Y_{i,j}&=&{\cal N}_{i,j}\left(\frac{B_1}{\sqrt{N}}\right)^{i-j},\;{\rm for}
\;i\geq j\nn Y_{i,j}&=&{\cal N}_{i,j}\left(\frac{B_1^{\dagger}}{\sqrt{N}}
\right)^{j-i},\;{\rm for}\;i\leq j,\nn
Y_{i,i}&=&{\cal N}_{i},\ea
where $1\leq i,j\leq N$.
The diagonal generators $Y_{i,i}$ mutually commute and are hermitian 
with respect to  $^*$ (not $^{\dagger}$), because the basis is now 
orthogonal (\ref{s}) in the dual Fock space (\ref{dualF}). Note that 
\bee\label{note}
Y_{i,1}Y_{1,j}=\left\{\begin{array}{ll}
Y_{i,j}\left({\cal N}_1+1\right)\cdots \left({\cal N}_1+j\right), & i\geq j\\
\left({\cal N}_1+1\right)\cdots \left({\cal N}_1+j\right)Y_{i,j}, & i\leq j .
\end{array}\right. \eeq
Hence, $Y_{i,j}$ generators can be written in terms of $2N-1$ $Y_{1,i}$ 
 generators and  their hermitian conjugates.
We find 
\baa \label{yalg}
\left[{\cal N}_1,Y_{i,j}\right]&=&-(i-j)Y_{i,j},\; \left[{\cal \br N},Y_{i,j}
\right]=(i-j)Y_{i,j},\nn
\left[H,Y_{i,j}\right]&=&\left[{\cal N},Y_{i,j}\right]=0,\;
\left[Y_{i,1},Y_{j,1}\right]=\left[Y_{1,i},Y_{1,j}\right]=0,\;\forall i,j,\nn
\left[Y_{i,1},Y_{1,j}\right]&=&\left[{\cal N}_{i,j}-\delta_{ij}{\cal N}_1\right]
B_1^{i-1}B_1^{\dagger (j-1)}+{\cal N}_{i,1}{\cal N}_{1,j}\left[
B_1^{i-1},B_1^{\dagger (j-1)}\right]\nn &=&
-\delta_{ij}{\cal N}_1g_{ii}({\cal N}_1)+Y_{i,1}Y_{1,j}f_{ij}({\cal N}_1)
,\;i>j,\ea
with $f$ and $g$ being some rational functions of ${\cal N}_1$.
The dynamical symmetry algebra
${\cal C}_N(\n)$ is  expressed in a  much simpler way through the above 
$\{Y_{i,j}\}$ generators than through the $\{X_{i,j}\}$ 
or $\{\tilde X_{s,s_s}\}$
 generators, but in all three
cases is still of the polynomial type.

The ${\cal C}_N(\n)$ algebras  are
equivalent (isomorphic) for all  $\n$ parameters larger than $-1/N$. 
Also, they
are 
equivalent (isomorphic) to the dynamical symmetry algebra ${\cal C}_N$
of the \h \ $H_2=H'-E_0=\sum_{k=1}^Nkb_k^{\dagger}b_k$,  where
$b_k$'s satisfy the standard bosonic commutation relation, 
with $b_1=B_1/\sqrt{N}$.
The generators of the corresponding symmetry algebra are
\bee 
Z_{i,j}=\left\{\begin{array}{ll}
b_i^{\dagger}b_j(b_1)^{i-j}, & i>j \\
b_i^{\dagger}b_j(b_1^{\dagger})^{j-i}, & i<j \\  
b_i^{\dagger}b_i, &i=j
\end{array}\right.\eeq
and $[H_2,Z_{i,j}]=0$. 
They act among degenerate states with fixed energy $E={\cal N},\; 
{\cal N}\in \Nb_0$, of the \h \ $H_2$.
The commutation relations for 
$Z_{i,j}$'s are the 
same as those for $Y_{i,j}$'s, see Eqs.(\ref{yalg}). This motivated 
us to look 
for a bosonic realization of the dynamical symmetry algebra.

{\bf Set 4}. Finally, we present the bosonic realization 
of the ${\cal C}_N(\n)$ algebra in 
terms of $\{{\cal J}_i^{\pm},{\cal J}_i^0\},\;i=2,\ldots N$ generators which 
represent a generalization of 
Eq.(\ref{su2}):
\baa\label{Jg}
{\cal J}_i^+&=&\frac{1}{\sqrt{i({\cal N}_1-1)\cdots({\cal N}_1-i+1)}}
b_1^{\dagger i}b_i, \nn
{\cal J}_i^-&=&b_i^{\dagger}b_1^i\frac{1}{\sqrt{i({\cal N}_1-1)\cdots
({\cal N}_1-i+1)}}=\left({\cal J}_i^+\right)^{\dagger},\nn
{\cal J}_i^0&=&\hlf\left(\frac{{\cal N}_1}{i}-b_i^{\dagger}b_i\right).\ea
One can express the generators $\{{\cal J}_i^{\pm},{\cal J}_i^0\}$ 
in terms of $\{A_k,A_k^{\dagger}\}$, using the expression for mapping,
Eq(\ref{gener}), for any $\n>-1/N$. The coefficients in the expansion depend 
on the parameter $\n$. 
The generators (\ref{Jg}) satisfy the following new algebra for 
every $i,j=2,\ldots, N$:
\baa\label{Jalg}
\left[{\cal J}_i^0,{\cal J}_j^{\pm}\right]&=&\pm\hlf\left(\frac{j}{i}
+\delta_{ij}\right){\cal J}_j^{\pm},\nn
{\cal J}_i^+{\cal J}_j^-&-&\left(\frac{\sqrt{{\cal N}_1({\cal N}_1-i+j)}}
{ {\cal N}_1+j}\right){\cal J}_j^-{\cal J}_i^+=\delta_{ij}\left(
\frac{{\cal N}_1}{i}\right),\nn
{\cal J}_i^+{\cal J}_j^+&-&\sqrt{\frac{{\cal N}_1-i}{{\cal N}_1-j}}
{\cal J}_j^+{\cal J}_i^+=0, \;
\left[{\cal J}_i^0,{\cal J}_j^0\right]=0.\ea
Note that the 
generators ${\cal J}_i^+$ and ${\cal J}_i^-$ are hermitian conjugate 
to each other, and that relations (\ref{Jalg}) do not depend on $\n$.
The ${\cal C}_N$ algebra contains $(N-1)$ ordinary $SU(2)$ subalgebras. 
Different $SU(2)$ subalgebras are connected in a non-linear way, 
namely relations (\ref{Jalg}) have non-linear (algebraic in ${\cal N}_1$)
deformations.
We demonstrate these features by the simple $N=3$ example:
\baa\label{Jalg3}
{\rm SU(2)}:\;&&
\left[{\cal J}_2^0,{\cal J}_2^{\pm}\right]=\pm{\cal J}_2^{\pm},\;
\left[{\cal J}_2^+,{\cal J}_2^-\right]=2{\cal J}_2^0,\nn
{\rm SU(2)}:\;&&
\left[{\cal J}_3^0,{\cal J}_3^{\pm}\right]=\pm{\cal J}_3^{\pm},\;
\left[{\cal J}_3^+,{\cal J}_3^-\right]=2{\cal J}_3^0,\nn
\left[{\cal J}_2^0,{\cal J}_3^{\pm}\right]&=&\pm\frac{3}{4}{\cal J}_3^{\pm},\;
\left[{\cal J}_3^0,{\cal J}_2^{\pm}\right]=\pm\frac{1}{3}{\cal J}_2^{\pm},\nn
{\cal J}_2^+{\cal J}_3^-&-&\frac{\sqrt{{\cal N}_1({\cal N}_1+1)}}{{\cal N}_1+3}
{\cal J}_3^-{\cal J}_2^+=0,\nn
{\cal J}_2^+{\cal J}_3^+&-&\sqrt{\frac{{\cal N}_1-2}{{\cal N}_1-3}}
{\cal J}_3^+{\cal J}_2^+=0.\nonumber\ea

The states with  fixed energy $E=
{\cal N}$ are given  by $\prod_{i=1}^Nb_i^{\dagger n_i}/\sqrt{n_i!}|0\rangle$
with $\sum_{i=1}^Nin_i={\cal N}$. These states are orthonormal and build an 
irreducible representation (IRREP) of the symmetry algebra ${\cal C}_N$. 

The dimension of the IRREP on the states with fixed energy ${\cal N}$
 of the ${\cal C}_N$ algebra can be 
obtained recursively in $N$:
$$D({\cal N},N)=\sum_{i=0}^{[{\cal N}/N]}
D({\cal N}-Ni,N-1).$$ For example, for $N=1$, 
$D({\cal N},1)=1$, i.e. for a single oscillator, there is no degeneracy;
for $N=2$, $D({\cal N},2)=\left[\frac{{\cal N}}{2}\right]+1$, ${\cal N}\geq 0$;
for $N=3$,
$$D({\cal N},3)=\sum_{i=0}^{\left[\frac{{\cal N}}{3}\right]}\left[\frac{{\cal N}
-3i}{2}\right]+\left[\frac{{\cal N}}{3}\right]+1. $$
More specifically, for ${\cal N}=6n+i$, for some integer $n$ and $i=0,\ldots,5$:
\baa
D(6n+i,3)&=&(3n+i)(n+1),\; {\rm for}\;i=1,\ldots,5,\nn
D(6n,3)&=&3n(n+1)+1.\nonumber\ea

Note that the dynamical symmetry  algebra of the 
\h \ $H_1=\sum_{k=1}^N
 b_k^{\dagger}b_k$ is $SU(N)$. The corresponding generators
are $b_i^{\dagger}b_j$, and $[H_1,b_i^{\dagger}b_j]=0$ for all $i$ and $j$.
The degenerate states  are $\prod_{i=1}^Nb_i^{\dagger n_i}/\sqrt{n_i!}|0\rangle$
 with $\sum_in_i={\cal N}$.
They form a totally symmetric IRREP of $SU(N)$ described by the Young diagram
with ${\cal N}$-boxes 
$\underbrace{\fbox{\rule{0mm}{0.4mm}}\framebox[4mm]{...}\fbox{\rule{0mm}{0.4
mm}}}_{\cal N}$ \cite{Sn}. 
The states of the same Fock space $\{\prod_i b_i^{\dagger n_i}/\sqrt{n_i!}
|0\rangle\}$ are differently arranged. With respect to  $H_1$,
degenerate states are organized into the $SU(N)$ 
$\underbrace{\fbox{\rule{0mm}{0.4mm}}\framebox[4mm]{...}\fbox{\rule{0mm}{0.4
mm}}}_{\cal N}$ symmetric  IRREP's. 
However, in respect to  $H_2$,
degenerate states build the 
${\cal N}$-IRREP of the ${\cal C}_N$ algebra. 
In this sense, one can (formally)
say that the non-linear  algebra ${\cal C}_N$ of the \h \ $H_2$
is related to the  boson 
realization of the $SU(N)$ algebra describing the dynamical 
symmetry of the \h \ $H_1$.

\section{The Chern-Simons matrix model}

\subsection{Introduction - The physical Fock space}

The Chern-Simons (CS) matrix model was obtained from the non-commutative $U(1)$ 
Chern-Simons theory, and it was conjectured that it described the quantum 
Hall fluid  \cite{suss}. The regularized version of the model, introduced in 
Ref.\cite{pol11}, is 
related to the Calogero model \cite{pol11,ks,jhep}. In Ref.\cite{jhep}
it was shown that the minimal algebra ${\cal A}_N$ of the observables was 
identical in the CS matrix theory and in the Calogero model. 
Using the results obtained in the  sections herebefore, we discuss the 
algebraic structure of the finite CS matrix theory in detail.
  
Let us start from the action proposed in Ref.\cite{pol11}:
\bee\label{cs1}
S=\int dt\frac{B}{2}{\rm Tr}\left\{\varepsilon_{ab}\left(\dot X_a+
i\left[A_0,X_a\right]\right)X_b+2\theta A_0-\omega X_a^2\right\}+
\Psi^{\dagger}\left(i\dot\Psi -A_0\Psi\right).\eeq
Here, $A_0$ and $X_a,\;a=1,2$, are $N\times N$ hermitian matrices and $\Psi$
is a
complex $N$-vector. The eigenvalues of the matrices $X_a$ represent the
coordinates of electrons and $A_0$ is a gauge field.
We  choose the gauge $A_0=0$ and impose the equation
of motion for $A_0$  as a constraint:
\bee \label{cs2}
-iB\left[X_1,X_2\right]+\Psi\Psi^{\dagger}=B\theta. \eeq
The trace part of Eq.(\ref{2}) gives
$\Psi^{\dagger}\Psi=NB\theta.$
Notice that the commutators have so far been classical matrix commutators.
After
quantization, the matrix elements of $X_a$ and the components of
$\Psi$ become operators satisfying the following commutation relations:
\baa \label{cs4}
\left[\Psi_i,\Psi_j^{\dagger}\right]&=&\delta_{ij},\nn
\left[\left(X_1\right)_{ij},\left(X_2\right)_{kl}\right]&=&\frac{i}{B}
\delta_{il}\delta_{jk}.\ea
It is convenient to introduce the operator $A=\sqrt{\frac{B}{2}}(X_1+iX_2)$ and
its hermitian conjugate $A^{\dagger}$ obeying the following commutation
relations:
\bee \label{cs5}
\left[\left(A\right)_{ij},\left(A^{\dagger}\right)_{kl}\right]=
\delta_{il}\delta_{jk} ,\;
\left[\left(A\right)_{ij},\left(A\right)_{kl}\right]=
\left[\left(A^{\dagger}\right)_{ij},\left(A^{\dagger}\right)_{kl}\right]=0.\eeq
Then, one can write the \h \ of the model at hand as
\bee \label{cs6}
H=\omega\left(\frac{N^2}{2}+{\rm Tr}(A^{\dagger}A)\right)=
\omega\left(\frac{N^2}{2}+{\cal N}_A\right),\eeq
${\cal N}_A$ being the total number operator associated with $A$'s.
Upon quantization, the constraints (\ref{cs2}) become the generators of unitary
transformations of both $X_a$ and $\Psi$.
The trace part of the constraint (\ref{cs2}) 
demands that (the l.h.s. being the number operator for
$\Psi$'s) $B\theta\equiv l$ be quantized to an integer.
The traceless part of the constraint (\ref{cs2}) demands that the
wave function be invariant under $SU(N)$ transformations, under which
$A$ transforms in the adjoint
and $\Psi$ in the
fundamental representation.
Note that as $A$ transforms in the reducible representation
$(N^2-1)+1$, with the singlet
$B_1={\rm Tr}\;A$, one can introduce a pure adjoint  representation as
$(\bar A)_{ij}=(A)_{ij}-\delta_{ij}B_1/N$. This slightly modifies the
commutator (\ref{cs5}), and completely decouples $B_1$ from the Fock space.
Physically, this correspondis to the separation of the center-of-mass coordinate
as it has been done for the Calogero model.  However, for the
sake of simplicity, this will not be done in this section.
So, one has to remember that when we compare the results for  the 
Calogero model with those for the CS matrix model, it is always up to 
the separation of the center-of-mass.

Energy eigenstates will be $SU(N)$ singlets, and 
explicit expressions for the wave functions were written in Ref.\cite{hell}:
\bee \label{cs8}
|\Phi\rangle=\prod_{i=1}^{N}({\rm Tr}A^{\dagger i})^{c_i}
C^{\dagger l}|0\rangle ,\eeq
where
$C^{\dagger}\equiv\varepsilon^{i_1\ldots i_N}\Psi^{\dagger}_{i_1}
(\Psi^{\dagger}A^{\dagger})_{i_2}\ldots (\Psi^{\dagger}A^{\dagger N-1})_{i_N}$,
and  $(A)_{ij}|0\rangle=\Psi_i|0\rangle=0$.

The system contains $N^2+N$
oscillators coupled by $N^2-1$ constraint equations
in the traceless part of Eq.(\ref{cs2}).  Effectively, we can describe the
system wih $N+1$ independent oscillators. Therefore,
the physical Fock space that consists  of all $SU(N)$-invariant states
can be spanned by $N+1$ algebraically independent
operators: 
\bee\label{defB}
B_n^{\dagger}\equiv {\rm Tr}A^{\dagger n},\;{\rm  with}\; n=1,2,\ldots,N,\eeq
and $C^{\dagger}$. Again, 
the operators $B_k^{\dagger}$
for $k>N$ can be expressed as a homogeneous polynomial of total order k in
$\{B_1^{\dagger},\ldots,B_N^{\dagger}\}$, with constant coefficients which
are common to all operators $A^{\dagger}$, see Appendix C.
We use the same letter to denote observables in  both models.
In the CS matrix model, $B_n={\rm Tr}\;A^n$ and in the Calogero model,
$B_n=\sum_ia_i^n$, but from the context it is clear what $B_n$
represent.
Since
\bee\label{vacuum}
{\rm Tr}\;A^kC^{\dagger l}|0\rangle\equiv
B_kC^{\dagger l}|0\rangle=0,\;\forall k,\forall l,\eeq
the state $C^{\dagger l}|0\rangle\equiv\va$ can be interpreted as a
ground state -  vacuum with respect to all operators $B_k$. Note
that the vacuum is not normalized to one, i.e., $\langle 0,l\va\neq 1$.
The whole physical Fock space can  be decomposed into  towers
(modules) built on the ground states with different $l$:
$$F_{\rm phys}^{\rm CS}=\sum_{l=0}^{\infty}F_{\rm phys}^{\rm CS}(l)=
\sum_{l=0}^{\infty}\{\prod B_k^{\dagger n_k}\va\}.$$
In Ref.\cite{jhep} it was shown that the states in the l-th
tower of the Chern-Simons
matrix model Fock space were identical to  the states of the Calogero
model with the interaction parameter $\n=l+1$.

\subsection{The ${\cal A}_N$ algebra, bosonic realization and
dynamical symmetry}

Using $[A_{ij},B_n^{\dagger}]=n(A^{\dagger n-1})_{ij}$, we find
a general expression for the commutators between observables:
\beq B
[B_m,B_n^{\dagger}]=n\sum_{r=0}^{m-1}{\rm Tr}\left(A^rA^{\dagger n-1}A^{m-r-1}
\right)=
m\sum_{s=0}^{n-1}{\rm Tr}\left(A^{\dagger s}A^{m-1}A^{\dagger n-s-1}\right)
.\eeq
One can normally order the r.h.s. of Eq.(\ref{B}) using the recurrent relation
\bee\label{no}
{\rm Tr}(A^rA^{\dagger n-1}A^{m-1})={\rm Tr}(A^{r-1}A^{\dagger n-1}A^m)+
\sum_{s=0}^{n-2}{\rm Tr}(A^{r-1}A^{\dagger s}){\rm Tr}(A^{\dagger n-s-2}
A^{m-1})
.\eeq
With the formal mapping
${\rm Tr}(A^rA^{\dagger s}A^k)\rightarrow \sum_ia_i^ra_i^{\dagger s}a_i^k$,
relation (\ref{B}) goes to the relation valid for the observables in the 
Calogero model.
Also, the
recurrent relation (\ref{no}) has its  counterpart in the Calogero model
with $\n=1$, with the same formal mapping.
The minimal algebra ${\cal A}_N$ including
only  observables of the type $B_n$ and $B_n^{\dagger}$,
defined by the following relations (including the corresponding
hermitian conjugate relations):
%\ba N
%\left[B_1,B_n^{\dagger}\right]&=&nB_{n-1}^{\dagger},\nn
%\left[B_2,B_n^{\dagger}\right]&=&2n {\rm Tr}A^{\dagger n-1}A+n\sum_{r=0}^{n-2}
%B_{r}^{\dagger}B_{n-r-2}^{\dagger},\nn
%\left[B_3,B_n^{\dagger}\right]&=&3n{\rm Tr}A^{\dagger n-1}A^2+
%3n\sum_{r=0}^{n-2}
%B_{r}^{\dagger}{\rm Tr}A^{\dagger n-r-2}A^{n-2}\nn
%&+&n\left[{n-1\choose 2}B_{n-3}
%^{\dagger}+\sum_{r=0}^{n-3}\sum_{s=0}^rB_s^{\dagger}B_{r-s}^{\dagger}B_{n-r-3}
%^{\dagger}\right].\ea
\beq D
[B_{i_1},[B_{i_2},[\ldots,[B_{i_n},B_n^{\dagger}]\ldots]]=n!\prod_{\alpha=1}^n
i_{\alpha}B_{I-n},\eeq
where $I=\sum_{\alpha=1}^ni_{\alpha}$ and $i_1,\ldots,i_n,n=1,2,\ldots, N$.
The identical successive commutators
 relations (\ref{D}) hold for the observables acting on the
$S_N$-symmetric Fock space of the Calogero model, see Eq.(\ref{d}), the 
only difference being that in the Calogero model we have separated the 
center-of-mass motion.
The {\it minimal} set of generalized vacuum conditions
needed to completely
define the representation of the algebra (\ref{D}) on the Fock space is
\ba V
B_2B_2^{\dagger}\va &=&2N(N+lN-l)\va,\nn
B_3B_3^{\dagger}\va &=&3N[N^2+1+l(N-1)(2N-1)+l^2(N-1)(N-2)]
\va=y\va,\nn
B_3B_3^{\dagger 2}\va &=&
54\{(l+1)B_1^{\dagger}B_2^{\dagger}+[N+(N-2)l+y/27]B_3^{\dagger}\}
\va.\ea
The generalized vacuum conditions for the
CS matrix
model (\ref{V})  are the same as those for the Calogero model with 
the interaction parameter $\n=l+1$ \cite{jhep}.

Using the results obtained in the Calogero model, we can construct 
a mapping from free Bose oscillators $\{b_i,b_i^{\dagger}\}$ 
to  $\{B_i,B_i^{\dagger}\}$. However, the coefficients in the 
expansion
$$ B_n^{\dagger}=\sum_k(\prod b^{\dagger})^{n+k}(\prod b)^k $$
depend on $l$, i. e. they are not the same for all towers in the 
physical Fock space.
On the other hand, we can construct a mapping that includes the 
$C$ operator and its bosonic conterpart  $c_0$:
$$ B_n^{\dagger}=\sum_{k,l}(\prod b^{\dagger})^{n+k}c_o^{\dagger l}c_0^l
(\prod b)^k, $$
$$C^{\dagger}=\sum_{k,l}(\prod b^{\dagger})^{k}c_o^{\dagger l+1}c_0^l
(\prod b)^k, $$
where
$$[b_i,b_j^{\dagger}]=\delta_{ij},\;[c_0,c_0^{\dagger}]=1,\;[b_i,c_0^{\dagger}]
=0.$$
The inverse mapping is given similarly as
$$b_n^{\dagger}=\sum_{k,l}(\prod B^{\dagger})^{n+k}C^{\dagger l}C^l
(\prod B)^k, $$
$$c_0^{\dagger}=\sum_{k,l}(\prod B^{\dagger})^{k}C^{\dagger l+1}C^l
(\prod B)^k. $$
As in the case of the  Calogero model, these mappings are not unique, but 
there exist a simple and natural choice of the boundary condition, 
see 
Eqs.(\ref{gfix}) and (\ref{gfixg}).
The mapping to bosons provides a natural orthogonal basis. 

In order
to describe the dynamical symmetry of the CS matrix model we
can use the same relations as those obtained for the Calogero model.
The generators of the symmetry algebra obtained from the Calogero model
act within a
fixed tower of states in $F_{\rm phys}^{\rm CS}$. However, one can introduce
an additional generator acting between different towers in order to
describe the larger dynamical symmetry connecting all degenerate states in the 
CS matrix model.

\subsection{The ${\cal B}_N^{\rm CS}$ algebra}

The problem of finding the underlying algebra of observables in the CS matrix 
model is more complicated than in the Calogero model. An additional problem 
is that in the CS matrix model there are more invariants of a given order. 
For example, let us consider the set of six observables of the fourth order
$\{\Tr(A^{\dagger 2}A^2),\Tr(AA^{\dagger 2}A),\Tr(A^{\dagger}A^2A^{\dagger}),
\Tr(A^2A^{\dagger 2}),\Tr(A^{\dagger}AA^{\dagger}A),\Tr(AA^{\dagger}
AA^{\dagger})
\}$. After performing the normal ordering (see Appendix C), 
we find that there are two 
independent,
normally 
ordered invariants  $\{\Tr(A^{\dagger 2}A^2),A_{ij}^{\dagger}A_{kl}^{\dagger}
A_{jk}A_{li}\}$, for $N\geq 4$. (For $N=3$ there is only one independent 
invariant of that order.) In the Calogero model, 
for $N\geq 4$ there is only one 
invariant of the fourth order $\sum a_i^{\dagger 2}a_i^2$. Hence, we cannot 
describe the ${\cal B}_N^{\rm CS}$  algebra in terms of the generators
$B_{m,n}=\Tr(A^{\dagger m}A^n)$ in a way analogous to the procedure in the 
Calogero model.

There are two approaches to the  construction of algebra of observables in the 
CS matrix model. Generally, these two approaches  result in different 
algebra, although in the Calogero model they produce the same algebra.
The first one is to write all possible invariants in 
$A^{\dagger}$ and $A$, and then to reduce them to the normally 
ordered ones. The main point is that this set of normally ordered inariants 
coincides with the 
set of all invariants in two $N\times N$ matrices $X$ and $Y$ of which the 
matrix elements are $c$ numbers. The problem thus reduces to finding 
the minimal number of algebraically independent invariants of two 
 $N\times N$ matrices.  This was solved explicitly for 
$N=2,3$ by Sibirski \cite{Sib} and for general $N$
by Donkin \cite{don}. One starts with the finite set of the 
algebraically independent invariants, for example, for $N=3$ with
$\{B_1,B_2,B_3,B_1^{\dagger},B_2^{\dagger},B_3^{\dagger}, \Tr(A^{\dagger}A),
\Tr(A^{\dagger 2}A),\Tr(A^{\dagger}A^2),\Tr(A^{\dagger 2}A^2),
:\Tr(A^{\dagger 2}A^2A^{\dagger}A):\}$, and calculates all possible 
commutators between them. The results are invariant operators that
can be expressed generally in terms of polynomials in algebraically 
independent invariants. Note that the basic set of independent invariants is not
 unique and can be chosen in different ways. However, each invariant can be 
uniquely expressed in terms of a given basic set of algebraically independent 
 invariants. Generally, one can express the commutator between any two 
observables as a polynomial in terms of basic generators. 
This approach is under investigation and will be reported 
separately.

The second approach is a generalization of our construction of the ${\cal A}_N$ 
algebra.
Namely, we start with a minimal number of generators $\{B_1,B_2,\ldots,B_
1^{\dagger},B_2^{\dagger},\ldots\}$ describing the complete physical Fock
space, and try to close the algebra under the Lie bracket (commutator) with a 
minimal number of additional generators. For finite $N$, the algebra is finite
and generally contains less generators compared with the first approach.
Since the result of the Lie-commutator of two generators is a polynomial 
in the basic generators, we call the algebra of this type finite 
polynomial Lie algebra.

For $N=2$, the ${\cal B}_2$ algebras for the Calogero model
 and the  CS matrix model 
are equivalent, but for $N\geq 3$ they differ. We demonstrate this 
by the $N=3$ example.
We start the construction of the ${\cal B}_3^{\rm CS}$ algebra with 
$\{B_1,B_2,B_3,B_1^{\dagger},B_2,^{\dagger}B_3^{\dagger}\}$ and 
four additional generators ${\cal O}_{1,1},{\cal O}_{1,2},{\cal O}_{2,1},
{\cal O}_{2,2}$ defined as
$$[B_i,B_j^{\dagger}]=ij{\cal O}_{i-1,j-1},\; i,j=2,3.$$
The operators $\{B_2,B_2^{\dagger},{\cal O}_{1,1}\}$ 
build the $su(1,1)$ algebra. This algebra is a subalgebra
of every polynomial Lie algebra ${\cal B}_N^{\rm CS}$. 
The operator ${\cal O}_{1,1}$, (the total number operator) satisfies the 
following relation:
$$[{\cal O}_{1,1},{\cal O}_{i,j}]=(i-j){\cal O}_{i,j}.$$
The  ${\cal B}_3^{\rm CS}$ algebra is defined by two types of 
commutation relations:
\bee\label{com1}[B_i,{\cal O}_{j,k}]=ij {\cal O}_{j-1,k+i-1},\;
[B_i^{\dagger},{\cal O}_{j,k}]=ik{\cal O}_{i+j-1,k-1},\eeq
and
\bee\label{com2}
[{\cal O}_{i+1,j+1},{\cal O}_{k+1,l+1}]=(jk-il){\cal O}_{i+k+1,j+l+1},\;
i,j,l,k=0,1.\eeq
The latter commutator (\ref{com2}) follows from  the former (\ref{com1}) and 
the Jacobi identities. 

It is important to note that the invariant of the sixth order $\Tr(A^{\dagger 
2}A^2A^{\dagger}A)$ does not participate in the above algebra! This means that 
although the above algebra ${\cal B}_3^{\rm CS}$ is the minimal algebra, 
closed under
commutation, it is not complete in the sense that we do not know 
the commutator between any two observables. The second important point is 
that the above algebra  ${\cal B}_3^{\rm CS}$ applies equally well to the 
three-body Calogero model. The only difference is that the invariant 
${\cal O}_{2,2}$ in the Calogero model can be expressed  in terms of the 
lower invariants, see Appendix A. 
However, when we reduce the set of generators for ${\cal B}_3^{\rm Cal}$ to 
nine, using the identity for ${\cal O}_{2,2}$, we obtain a different 
algebra with different commutation relations.
The above construction can be generalized for any finite $N$. The generators 
are of the form $B_n$, $B_m^{\dagger}$, $ [B_n,B_m^{\dagger}]$, 
$[B_k,[B_n,B_m^{\dagger}]]$, etc. The number of generators is finite, with an
upper limit of $2N^2$ generators. Of course, the lower limit of the number of 
generators is the number of generators 
of ${\cal B}_N^{\rm Cal}$ in the Calogero model - $1/2(N+4)(N-1)$.
Using Jacobi
identities the commutator of any two generators can 
be expressed as a combination of successive commutators. 
Some of them are generators of the algebra, and the others can be 
expressed as polynomials of generators. Here we give some simple, 
general results:
\baa
&&\frac{[B_{i_{n-1}},\ldots,[B_i,B_n^{\dagger}]\cdots]}{\left(\prod_{\alpha
=1}^{n-1}i_{\alpha}\right) n!}=\frac{[B_{I'-n+2},B_2^{\dagger}]}{2(I'-n+2)},\;
I'=\sum_{\alpha=1}^{n-1}i_{\alpha},\nn
&& \Tr(A^{\dagger r}AA^{\dagger s}+A^{\dagger s}AA^{\dagger r})=
\Tr(A^{\dagger r+s}A+AA^{\dagger r+s}).\ea

The authors of Ref.\cite{poly-mx} 
considered the infinite 
set of all successive commutators of the type $[g_{i_1},[g_{i_2},\ldots,
[g_{i_n},g_{i_{n+1}}]\cdots]$, where $g_i$ is either $B_i$ or $B_i^{\dagger}$,
$i$ natural number. This set was closed under commutation, which followed from the
Jacobi identities, and all generators were "mirror" symmetric.  However, they did
not know how the algebra ${\cal G}$ looked like explicitly,  and gave a step
by step construction. As their main results they presented a Table with numbers
of linearly independent invariants of a given type $(n,m)$, but they completely 
omitted the discussion of non-linear identities between invariants that are
crucial for the finite number of degrees of freedom. The
conclusion of Ref.\cite{poly-mx} was that the algebras  ${\cal G}$ 
for the Calogero model
and  for the matrix model were the same.
Although this is true, the actual
models are different, as can be seen from the
algebraic structure obtained for a fixed number of
particles $N$.

\section{Conclusion}

The minimal algebra ${\cal A}_N$ of invariant operators $A_i$ 
defined by the successive
commutation relation  and the generalized vacuum conditions completely
define the action of the operators $A_i$ on the states in the physical Fock space.
The general structure of the ${\cal A}_N$ algebra can be viewed as a
generalization of triple operator algebras \cite{trip} to the $(N+1)$-tuple
operator algebra. For $N=2$, the ${\cal A}_2(\n)$ algebra is just
$[A_2,[A_2,A_2^{\dagger}]]=8A_2$ for a single oscillator $A_2=(a_1-a_2)^2/2$
describing the relative motion in the two-body problem.

We have constructed an orthogonal basis $\A_i$ in the dual Fock space.
The operators $\A_i$ and $A_j^{\dagger}$ are
conjugate to each other with respect to the new scalar product
that they induce  on the dual Fock space.
This construction differs from the construction proposed in
Ref.\cite{ind} where conjugated operators with different indices do not commute.
Furthermore, we have shown
that there exists a mapping from ordinary Bose oscillators to operators
$A_k,A_k^{\dagger}$, and its inverse, and in this way we have 
obtained a natural
orthogonal basis for the symmetric Fock space.

Since this algebraic structure is identical in the Calogero model with that in
the CS matrix model, all above-mentioned  results aplly equally to both models.

It has  been shown \cite{ks,jhep} that the states in the
$l-$th tower of the CS matrix model Fock space  are equivalent to
the states in the physical
Fock space of the Calogero model with the interaction parameter $\nu=l+1$.
Therefore, in order
to describe the dynamical symmetry of the CS matrix model, we
can use the same relations as those obtained for the Calogero model.

The authors of Ref.\cite{poly-mx} claim 
that the infinite algebra ${\cal G}$, analyzed in their paper, is common
to both, the Calogero model  and the CS matrix model.
In fact, this algebra is common to all systems of infinitely many identical
 particles.  However, the analysis of the finite CS matrix model proposed in 
Ref.\cite{pol11} requires finite algebras of observables.
We have shown that owing to the trace
identities between observables for finite $N$, the effective, minimal,
finite ${\cal B}_N$ algebras of observables for  the Calogero model  and the 
CS matrix model are quite different.
The ${\cal B}_N^{\rm CS}$ algebra contains more algebraically independent 
observables than 
${\cal B}_N^{\rm Cal}$. Also, in the CS matrix model one has to be careful
when constructing the ${\cal B}_N^{\rm CS}$ algebra. 
There are two different methods 
of constructing the algebra and they lead to different results, 
as we have demonstrated 
by the $N=3$ example.
Moreover, for the 
Calogero model, ${\cal B}_N^{\rm Cal}$ can be connected to $W$ algebras, 
whereas for the CS matrix model  it is unclear whether ${\cal B}_N^{\rm CS}$ 
represents a generalization of the $W$ algebra.

Acknowledgment

We would like to thank V. Romanovsky and D. Svrtan for useful
discussions.
This work was supported by the Ministry of Science and Technology of the
Republic of Croatia under contract No. 00980103.

\appendix

\section*{Appendix A}

It is known \cite{Mac,MF}
that the sum of powers $B_n=\sum_{i=1}^Na_i^n$ for $n>N$ can be 
expressed in terms of $B_k,\;1\leq k\leq N$ in the form
$B_n=\sum\left(\prod B\right)^n\;.$
In the preceding sections we have 
claimed that there exist similar relations for 
 the elements of the  algebras $A_N(\n)$ and $B_N(\n)$ of the form
\bee\label{a1}
A_n=\sum\left(\prod A\right)^n\;,n>N,\eeq
and more generally,
\bee\label{a2}
\br B_{n,m}=\sum\left(\prod\br B_{n_{\alpha},m_{\alpha}}\right),\;n+m>N,\eeq
where $\sum n_{\alpha}\leq n,\;\sum m_{\alpha}\leq m$.
Here we give a method for calculating the coefficients in the 
identities (\ref{a1}) 
and (\ref{a2}). 

For  fixed $N$, let us denote
\bee\label{a3}
A_n=\sum_{i=1}^{N-1}\br a_i^n+(-)^n\left(\sum_{i=1}^{N-1}\br a_i\right)^n,\eeq
where $\br a_1,\ldots, \br a_{N-1}$ are independent operators. Since these 
operators mutually commute, we treat the above identity as an identity in the 
ring of polynomials in real variables $a_i\in\Rb$. In order to calculate the 
coefficients in Eq.(\ref{a1}), we construct a set of linear equations 
inserting some points $(a_1,a_2,\ldots,a_{N-1})$ in  Eq.(\ref{a1}). For 
example, the point $(\underbrace{1,k,0,\ldots,0}_{N-1})$ gives
$A_n=1+k^n+(-)^n(1+k)^n$ for $A_n$ in Eq.(\ref{a3}).
For $N=2$, one easily finds
$$A_{2k}=\frac{1}{2^{k-1}}A_2^k,\;A_{2k+1}=0,\;k>1.$$
We also list some results for $N=3$:
\baa
A_4&=&\hlf A_2^2,\nn
A_5&=&\frac{5}{6}A_2A_3,\nn
A_6&=&\frac{1}{4}A_2^3+\frac{1}{3}A_3^2,\nonumber\ea
and for $N=4$:
\baa
A_5&=&\frac{5}{6}A_2A_3,\nn
A_6&=&\frac{3}{4}A_2A_4-\frac{1}{8}A_2^3+\frac{1}{3}A_3^2,\nn
A_7&=&\frac{7}{12}A_3A_4+\frac{7}{24}A_2^2A_3 .\nonumber\ea

To calculate the coefficients in Eq.(\ref{a2}), we proceed in two steps. 
First, we find the coefficients in the following relation:
\bee\label{a4}
\br B_{n,m}=\sum:\prod\br B_{n_{\alpha},m_{\alpha}}:,\eeq
where $\sum n_{\alpha}=n,\;\sum m_{\alpha}=m$, and $:{\cal O}:$ denotes normal 
ordering of the operator ${\cal O}$,  $\br a_i^{\dagger}$ on left and $\br a_j$ 
on the right. Since by definition
$:\,a_ia_j^{\dagger}\,:\;=\;:\,a_j^{\dagger}a_i\,:$, 
we consider the identity (\ref{a4}) as
an identity in the ring of polynomials in two sets of commuting, real 
variables.
Then we construct a set of linear equations inserting some points in relations 
(\ref{a2}) and (\ref{a4}).
For example, for points
$\{\br a_1^{\dagger},\ldots,\br a_{N-1}^{\dagger}\}
=\{1,k,0,\ldots,0\}\;{\rm and}\;
\{\br a_1,\ldots,\br a_{N-1}\}=\{l,0,\ldots,0,1\} $
the corresponding invariants are:
\baa
\br B_{n,m}&=&l+(-)^{m+n}(1+k)^n(1+l)^m,\;m>0,\forall n,\nn
A_n^{\dagger}&=&1+k^n+(-)^n(1+k)^n,\forall n,\nn
A_n&=&1+l^n+(-)^n(1+k)^n,\forall n.\nonumber\ea
For the $N=2$ case, we find
\baa
&&\br B_{n,m}=0,\; {\rm for}\;n+m={\rm odd},\nn
&&\br B_{2k,2l}=\frac{1}{2^{k+l-1}}A_2^{\dagger k}A_2^l,
\br B_{2k+1,2l+1}
=\frac{1}{2^{k+l-1}}A_2^{\dagger k}\br B_{1,1}A_2^l.\nonumber\ea
We also have a set of relations for $N=3$:
\baa
\br B_{3,1}&=&\hlf A_2^{\dagger}\br B_{1,1},\nn
\br B_{2,2}&=&\frac{1}{3}:\br B_{1,1}^2:+\frac{1}{6}A_2^{\dagger}A_2,\nn
\br B_{4,1}&=&\frac{1}{3}A_3^{\dagger}\br B_{1,1}+\hlf A_2^{\dagger}\br B_{2,1},\nn
\br B_{3,2}&=&\frac{7}{12}A_3^{\dagger}A_2+\frac{3}{4}A_2^{\dagger}\br B_{1,2}
-\hlf :\br B_{2,1}\br B_{1,1}:. \nonumber\ea
In the 
four-particle case, the expressions for $\br B_{4,1}$ and $\br B_{3,2}$ are 
the same as in the three-particle case, and we 
present results for $B_{5,1}$, $\br B_{4,2}$ and 
$\br B_{3,3}$:
\baa
\br B_{5,1}&=&\frac{1}{3}A_3^{\dagger}\br  B_{2,1}+\hlf A_2^{\dagger}B_{3,1},\nn
\br B_{4,2}&=&\frac{1}{4}A_4^{\dagger}A_2-\frac{1}{8}A_2^{\dagger 2}A_2+\hlf 
A_2^{\dagger}\br B_{2,2}+\frac{1}{3}A_3^{\dagger}\br B_{1,2},\nn
\br B_{3,3}&=&-\frac{1}{24}A_3^{\dagger}A_3+\frac{3}{4}:\br B_{2,2}\br B_{1,1}:
-\frac{1}{8}:\br B_{1,1}^3:+\frac{3}{8}:\br B_{2,1}\br B_{1,2}:. \nonumber\ea
Note that identities for  $\br B_{n,m}$  transform into identities for $A_{n+m}$
 after identification $\br a_i^{\dagger}=\br a_i$.
  
In the second step we express the normally ordered
products in terms of the elements 
of the $\br B_N(\n)$ algebra:
\bee :\prod\br B_{n_{\alpha},m_{\alpha}}:\;=\sum\left(\prod\br 
B_{n'_{\alpha},m'_{\alpha}}\right), \;\sum n'_{\alpha}\leq n_{\alpha},\;
\sum m'_{\alpha}\leq m_{\alpha},\eeq
using the commutation relations (\ref{bar6}).
We present few examples:
\baa
:\br B_{1,1}^2:&=&\br B_{1,1}(\br B_{1,1}-1+N\n),\nn
:\br B_{2,1}\br B_{1,1}:&=&\br B_{2,1}(\br B_{1,1}-1+N\n),\nn
:\br B_{2,1}\br B_{1,2}:&=&\br B_{2,1}\br B_{1,2}-(1-N\n)\left(\br B_{2,2}
-\frac{1}{N}A_2^{\dagger}A_2 \right),\nn
:\br B_{2,2}\br B_{1,1}:&=&\br B_{2,2}(\br B_{1,1}-2+N\n)-\n A_2^{\dagger}
A_2,\nn
:\br B_{1,1}^3:&=&\br B_{1,1}(\br B_{1,1}-1+N\n)(\br B_{1,1}-2+2N\n-2\n)
-\n(A_2^{\dagger}A_2+N\br B_{2,2}).\nonumber\ea
Finally, we have
\baa
\br B_{2,2}&=&\frac{1}{3}\br B_{1,1}^2-\frac{1}{3}\br B_{1,1}(1-N\n)
+\frac{1}{6}A_2^{\dagger}A_2,\; N\leq 3,\nn
\br B_{3,2}&=&\frac{7}{12}A_3^{\dagger}A_2+\frac{3}{4}A_2^{\dagger}\br B_{1,2}
-\hlf\br B_{2,1}\br B_{1,1}+\hlf\br B_{2,1}(1-N\n),\; N\leq 4,\nn
\br B_{3,3}&=&-\frac{1}{24}A_3^{\dagger}A_3-\frac{1}{8}\br B_{1,1}(\br B_{1,1}
-1+N\n)(\br B_{1,1}-2+2N\n-2\n)+\frac{3}{4}\br B_{2,2}\br B_{1,1}
\nn &+&\frac{3}{8}\br B_{2,1}\br B_{1,2}
-\frac{5}{4}\br B_{2,2}\left(\frac{3}{2}-N\n
\right)+A_2^{\dagger}A_2\left(\frac{3}{8N}-\n\right),\;N\leq 5 .
\nonumber\ea

\section*{Appendix B}

In this appendix we perform some explicit calculations for the ${\cal A}_3(\n)$ 
algebra. 
The minimal set of relations
which define the ${\cal A}_{3}$ algebra of operators $\{A_2,A_3,
A_2^{\dagger},A_3^{\dagger},{\cal \br N}\}$ is
\ba e
&&[A_i[A_2,A_2^{\dagger}]]=4iA_i,\nn
&&[A_3,[A_3,A_2^{\dagger}]]=3A_2^2,\nn
&&[A_i,[A_2,[A_3,A_3^{\dagger}]]]=6iA_iA_2,\;i=2,3,\nn
&&[A_2,[A_2,[A_2,A_3^{\dagger}]]]=48 A_3,\nn
&&[A_3,[A_3,[A_3,A_3^{\dagger}]]]=54A_3^2-\frac{9}{2}A_2^3.
\ea
For a given representation with fixed parameter $\n$, we also need
generalized vacuum conditions:
\baa\label{gvc}
&&A_2|0\rangle=A_3|0\rangle =A_2A_3^{\dagger}|0\rangle=
A_3A_2^{\dagger}|0\rangle=A_3A_2^{\dagger 2}|0\rangle=0,\nn
&&A_2A_2^{\dagger}|0\rangle =4(1+3\n)|0\rangle,\;
A_3A_3^{\dagger}|0\rangle=2(1+3\n)(2+3\n)|0\rangle,\nn
&&A_3A_2^{\dagger}A_3^{\dagger}|0\rangle=2(2+3\n)(4+3\n)A_2^{\dagger}
|0\rangle,\;
A_3A_3^{\dagger 2}|0\rangle=2(2+3\n)(11+6\n)A_3^{\dagger}|0\rangle.\ea
The algebra (\ref{e}), with the vacuum conditions (\ref{gvc}),
has a unique representation on $F_{\rm symm}$. Using
Eqs.(\ref{e}) and (\ref{gvc}) one finds
the action of the operators $A_2$ and $A_3$ on any state in the Fock space:
\ba f
&&A_2|n_2,n_3\rangle
=3{n_3\choose 2}|n_2+2,n_3-2\rangle
+4n_2(3n_3+n_2+3\n)|n_2-1,n_3\rangle,\nn
&&A_3|n_2,n_3\rangle=2\left(n_3(2+3\n)(1+3\n+3n_2)+9{n_3\choose 2}(2+3\n+n_2)
+27{n_3\choose 3}+6n_3{n_2\choose 2}\right)\times\nn
&&\times |n_2,n_3-1\rangle +
48{n_2\choose 3}|n_2-3,n_3+1\rangle
-\frac{81}{2}{n_3\choose 3}|n_2+3,n_3-3\rangle.
\ea
The ket  $|n_2,n_3\rangle$ denotes the state
$A_2^{\dagger n_2}A_3^{\dagger n_3}|0\rangle$.

Here we demonstrate how to construct 
an orthogonal basis in the dual Fock space defined by (\ref{dualF}).
We write the general expression (\ref{gen3}), up to $k+n\leq 5$:
\baa\label{b11}
\A_2&=&g_2A_2+g_{22}A_2^{\dagger}A_2^2+g_{23}A_3^{\dagger}A_2A_3
+\cdots\nn
\A_3&=&g_3A_3+g_{32}A_2^{\dagger}A_2A_3+\cdots ,\ea
with unknown coefficients g.
To calculate five unknown
coefficients   in (\ref{b11}), we need the following relations:
\baa\label{b3}
A_2A_2^{\dagger}|0\rangle&=&4(1+3\n)|0\rangle,\nn
A_3A_3^{\dagger}|0\rangle&=&2(1+3\n)(2+3\n)|0\rangle,\nn
A_2A_2^{\dagger 2}|0\rangle&=&8(2+3\n)A_2^{\dagger}|0\rangle,\nn
A_2A_2^{\dagger}A_3^{\dagger}|0\rangle&=&4(4+3\n)A_3^{\dagger}|0\rangle,\nn
A_3A_2^{\dagger}A_3^{\dagger}|0\rangle&=&2(2+3\n)(4+3\n)A_2^{\dagger}|0\rangle
.\ea
Next, we apply  Eqs.(\ref{b11})
to the
 states in the Fock space, and using (\ref{b3})
we obtain
\baa
\A_2&=&\frac{1}{4(1+3\n)}A_2-\frac{1}{16(1+3\n)^2(2+3\n)}A_2^{\dagger}A_2^2-
\frac{3}{8(1+3\n)^2(2+3\n)(4+3\n)}A_3^{\dagger}A_3A_2+\cdots,
\nn
\A_3&=&\frac{1}{2(1+3\n)(2+3\n)}A_3-\frac{3}{8(1+3\n)^2(2+3\n)(4+3\n)}
A_2^{\dagger}A_2A_3+\cdots
\nonumber .\ea

Next, we calculate the coefficients in the expressions (\ref{gener}), 
up to $k+n\leq 5$,
thus finding the mapping from the Bose oscillators $\{b_i,b_i^{\dagger}\}$ to 
the operators $\{A_i, A_i^{\dagger}\}$, and vice versa.
We write the general expression
\baa\label{b1}
A_2&=&f_2b_2+f_{22}b_2^{\dagger}b_2^2+f_{23}b_3^{\dagger}b_2b_3
+\cdots\nn
A_3&=&f_3b_3+f_{32}b_2^{\dagger}b_2b_3+\cdots ,\ea
with unknown coefficients f.
Inserting the expansion (\ref{b1}) into Eqs.(\ref{b3})  and solving for f's,
we obtain
\baa\label{fA}
A_2&=& 2\sqrt{1+3\n}\;b_2+2\left(\sqrt{2+3\n}-\sqrt{1+3\n}\right)
b_2^{\dagger}
b_2^2+2\left(\sqrt{4+3\n}-\sqrt{1+3\n}\right)b_3^{\dagger}b_2b_3+\cdots,\nn
A_3&=&\sqrt{2(1+3\n)(2+3\n)}\; b_3+
\left(\sqrt{2(4+3\n)(2+3\n)}-\sqrt{2(1+3\n)(2+3\n)}\right)b_2^{\dagger}b_2b_3+
\cdots,
\ea
and similarly for hermitian conjugates.

The norms in the $\{A_2^{\dagger n_2}A_3^{\dagger n_3}|0\rangle\}$ Fock space
are positive if $\n>-1/3$ (see Eq.(\ref{norma})), 
and we look for inverse mapping. 
The general expressions are
\baa\label{bgen}
b_2&=&f_2'A_2+f_{22}'A_2^{\dagger}A_2^2+f_{23}'A_3^{\dagger}A_2A_3
+\cdots\nn
b_3&=&f'_3A_3+f_{32}'A_2^{\dagger}A_2A_3+\cdots.\ea
Inserting these relations into 
$$b_ib_j^{\dagger}|0\rangle=\delta_{ij}|0\rangle,\;b_ib_i^{\dagger}b_j^{\dagger}
|0\rangle=(1+\delta_{ij})b_j^{\dagger}|0\rangle,$$
and solving for unknown coefficients, we find
\baa
b_2&=&\frac{A_2}{2\sqrt{1+3\n}}+\frac{A_2^{\dagger}A_2^2}{8(1+3\n)^{3/2}}
\left(
\sqrt{\frac{1+3\n}
{2+3\n}}-1\right)\nn &+&\frac{A_3^{\dagger}A_2A_3}{4(1+3\n)^{3/2}(2+3\n)}\left(
\sqrt{\frac{1+3\n}{4+3\n}}-1\right)+\cdots,\nn
b_3&=&\frac{A_3}{\sqrt{2(1+3\n)(2+3\n)}}+\frac{A_2^{\dagger}A_2A_3}
{4\sqrt{2(1+3\n)^3(2+3\n)}}
\left(\sqrt{\frac{1+3\n}{4+3\n}}-1\right)+\cdots.\ea

After determining the mapping, we are in a position to construct the orthogonal 
states. The first few of these states are as follows:
\baa\label{orthstates}
&&b_2^{\dagger}|0\rangle=\frac{1}{2\sqrt{1+3\n}}A_2^{\dagger}|0\rangle,\;
b_3^{\dagger}|0\rangle=\frac{1}{\sqrt{2(1+3\n)(2+3\n)}}A_3^{\dagger}|0\rangle
,\nn
&&b_2^{\dagger 2}|0\rangle=\frac{1}{4\sqrt{(1+3\n)(2+3\n)}}A_2^{\dagger 2}
|0\rangle,\nn
&&b_2^{\dagger}b_3^{\dagger}|0\rangle=\frac{1}
{2\sqrt{2(1+3\n)(2+3\n)(4+3\n)}}A_2^{\dagger}A_3^{\dagger}|0\rangle,\nn
&&b_2^{\dagger 3}|0\rangle=\frac{1}{8\sqrt{3(1+3\n)(2+3\n)(1+\n)}}
A_2^{\dagger 3}|0\rangle ,\nn
&&b_3^{\dagger 2}|0\rangle=\alpha\left(
-12(1+\n)A_3^{\dagger 2}|0\rangle+A_2^{\dagger 3}|0\rangle\right),
\nn &&\alpha^{-1}=12\sqrt{2(1+\n)(1+3\n)(2+3\n)
[(1+\n)(2+3\n)(11+6\n)-2]}.\ea

\section*{Appendix C}

We propose a method for constructing  algebraic relations between 
observables
in the Chern-Simons matrix model.
The starting point is the Cayley-Hamilton theorem. For a $N\times N$ matrix 
$A$, it expresses $A^N$ as a linear function of lower powers of $A$, with 
coefficients which are symmetric functions of eigenvalues of $A$:
\bee\label{CHT}
A^N=e_1A^{N-1}-e_2A^{N-2}+\ldots(-)^{N-1}e_N\cdot\JM,\eeq
where $e_r$ is the $r$th elementary symmetric function \cite{Mac}
of eigenvalues of $A$.
The trace of Eq.(\ref{CHT}) can be written in the following form:
\bee\label{trN}
p_N=\sum_{i=1}^N(-)^{i-1}e_ip_{N-i}, \eeq
where $p_r$ is the $r$th power sum \cite{Mac} of eigenvalues of $A$.
Using Eqs. (\ref{CHT}) and (\ref{trN}) we can  generate 
algebraic relations between observables, for  fixed $N$. We will 
demonstrate the method for the $N=2$ and $N=3$ cases.

For $N=2$, the Cayley-Hamilton theorem gives
\bee\label{ch2}
A^2=({\rm Tr}\,A)A-{\rm det}A\cdot\JM=B_1 A-{\rm det}A\cdot \JM,\eeq
and the trace of (\ref{ch2}) gives
\bee\label{tr2}
B_2=B_1^2-2{\rm det}A.\eeq  
First, we generate some algebraic relations expressing observable the $B_n$, 
defined in (\ref{defB}), for $n\geq 3$, using 
only $B_1$ and $B_2$. One multiplies Eq.(\ref{ch2}) by $A$, takes a trace and 
expresses ${\rm det}A$ using Eq.(\ref{tr2}), to obtain:
\bee\label{tr3}
B_3=\frac{3}{2}B_2B_1-\hlf B_1^3.\eeq
The expression for $B_4$ is obtained along the same lines, but we 
also have to 
use the relation (\ref{tr3}):
\bee\label{tr4}
B_4=\hlf B_2^2+B_2B_1^2-\hlf B_1^4.\eeq
Recursively, we can generete all algebraic relations expressing $B_n$, 
$n\geq 3$, 
using only $B_1$ and $B_2$. Of course, when we put $B_1=0$, we obtain results 
valid for  $A_j$ operators in the Calogero model (see Appendix A).

For the $N=3$, case we present a few relations obtained along the same lines.
The Cayley-Hamilton theorem gives
\bee\label{ch3}
A^3=(\Tr A)A^2+\hlf[\Tr A^2-(\Tr A)^2]A+{\rm det}A\cdot\JM,\eeq
and from the trace of Eq.(\ref{ch3}) we obtain
\bee\label{ctr3}
{\rm det}A=\frac{1}{3}B_3-\frac{1}{2}B_1B_2+\frac{1}{6}B_1^3.\eeq
Using Eqs.(\ref{ch3}) and (\ref{ctr3}) one easily obtains
\baa\label{n31}
B_4&=&\hlf B_2^2-B_2B_1^2+\frac{4}{3}B_3B_1+\frac{1}{6}B_1^4,\nn
B_5&=&\frac{5}{6}B_2B_3+\frac{5}{6}B_3B_1^2-\frac{5}{6}B_2B_1^3+\frac{1}{6}B_1^5
,\nn
B_6&=&\frac{1}{4}B_2^3+\frac{1}{3}B_3^2+B_3B_2B_1+\frac{1}{3}B_3B_1^3-
\frac{3}{4}B_2^2B_1^2-\frac{1}{4}B_2B_1^4+\frac{1}{12}B_1^6.\ea

Next, we turn to
the elements of the algebra ${\cal B}^{\rm CS}_N$.  
These observables are all of the type
$\Tr (A^{\dagger n_1}A^{m_1}\cdots A^{\dagger n_k}A^{m_k})$, where 
$m_i,n_i\in \{1,2,\ldots,(N-1)\}$, and $k<N$. They are not all algebraically
indepedent, and in the following we give a method for constructing the 
relations between them.

We are 
dealing with matrices whose matrix elements are operators, so we have to take 
care of ordering. It is important to observe that  
the relations between 
normally ordered invariants are identical to relations between invariants
in two $N\times N$ matrices whose matrix 
elements are $c$-numbers. Therefore, our first step is to 
reduce observables to normally ordered ones. For example,
\baa\label{nor}
&&\Tr(AA^{\dagger n})=\Tr(A^{\dagger n}A)+\sum_{s=0}^{n-1}\Tr A^{\dagger s}
\Tr A^{\dagger n-s-1},\nn
&&\Tr(AA^{\dagger}AA^{\dagger})=:\Tr(AA^{\dagger}AA^{\dagger}):
+3N\Tr(A^{\dagger}A)+N^3,\nn
&&\Tr(A^{\dagger}AA^{\dagger}A)=:\Tr(A^{\dagger}AA^{\dagger}A):
+\Tr(A^{\dagger}A),\nn
&&\Tr(A^2A^{\dagger 2})=\Tr(A^{\dagger 2}A^2)+3N\Tr(A^{\dagger}A)
+2\Tr A^{\dagger}\Tr A+N(N^2+1),\nn
&&\Tr(AA^{\dagger 2}A)=\Tr(A^{\dagger 2}A^2)+\Tr A^{\dagger}\Tr A+
N\Tr(A^{\dagger}A),\nn
&&\Tr(A^{\dagger}A^2A^{\dagger})=\Tr(A^{\dagger 2}A^2)+
\Tr A^{\dagger}\Tr A+N\Tr(A^{\dagger}A),\nn
&&\Tr(A^{\dagger 2}AA^{\dagger})=\Tr(A^{\dagger 3}A)+N\Tr A^{\dagger 2},\nn
&&\Tr(A^{\dagger}AA^{\dagger 2})=\Tr(A^{\dagger 3}A)+N\Tr A^{\dagger 2}+
(\Tr A^{\dagger})^2.
\ea
Now, we can  construct the relations connecting normally orderd 
observables starting from the expressions for $B_k\equiv \Tr(A^k)$.
For example, in the $N=3$ case, we use the first relation in Eq.(\ref{n31}) to write
\baa\label{exa}
:\Tr(A^{\dagger}+A)^4:&=&\hlf:(\Tr(A^{\dagger}+A)^2)^2:-:\Tr(A^{\dagger}+A)^2:
:(\Tr(A^{\dagger}+A))^2:\nn &+&\frac{4}{3}:\Tr(A^{\dagger}+A)^3\Tr(A^{\dagger}
+A):+
\frac{1}{6}:(\Tr(A^{\dagger}+A)):^4.\ea
This relation gives us an identity expressing observables of order $(3,1)$ as
functions of observables of lower order. An observable of order $(m,n)$ is 
any observable of the type $\Tr(AA^{\dagger }\cdots A\cdots)$ 
with $m$ $A^{\dagger}$
 and $n$ $A$ matrices in the trace. 
We take $:\Tr(A^{\dagger 3}A+A^2A^{\dagger}A+
AA^{\dagger}A^2+AA^{\dagger 3}):$ on the l.h.s of Eq.(\ref{exa}) and 
we pick up all terms of the same order in $A^{\dagger}$ and $A$ on the r.h.s.
Thus, we obtain the following result:
\baa\label{res31}
\Tr(A^{\dagger 3}A)&=&\hlf \Tr A^{\dagger 2}\Tr(A^{\dagger}A)-\hlf
\Tr A^{\dagger 2}\Tr A^{\dagger}\Tr A-\hlf (\Tr A^{\dagger})^2
\Tr(A^{\dagger}A)\nn &+&\frac{1}{6}(\Tr A^{\dagger})^3\Tr A+
\Tr A^{\dagger}\Tr( A^{\dagger 2}A)+\frac{1}{3}\Tr A^{\dagger 3}\Tr A,\ea
Using (\ref{res31}) and  relations given  in (\ref{nor}) we can 
express all observables of order  $(3,1)$ as 
functions of observables of lower order.
From relation (\ref{exa}) we also obtain a relation for observables 
of order  $(2,2)$ - we simply pick the terms of order $(2,2)$ on 
both sides of Eq.(\ref{exa}):
\baa
&&2\Tr( A^{\dagger 2}A^2)+:\Tr(A^{\dagger}AA^{\dagger}A):=
:(\Tr(A^{\dagger}A))^2:+\hlf \Tr A^{\dagger 2}\Tr A^2-
2\Tr A^{\dagger}\Tr( A^{\dagger}A)\Tr A \nn && -\hlf \Tr A^{\dagger 2}
(\Tr A)^2-\hlf
(\Tr A^{\dagger})^2\Tr A^2+\hlf (\Tr A^{\dagger})^2(\Tr A)^2+2
\Tr( A^{\dagger 2}A)\Tr A+2\Tr A^{\dagger}\Tr(A^{\dagger}A^2).\nonumber\ea

To obtain all relations among the invariants one uses the Cayley-Hamilton theorem
for matrix $C=A^{\dagger}+\lambda A$. For $k>N$ we can write 
$\Tr C^k$ as a  polynomial in $\Tr C,\ldots,\Tr C^N$, and project 
terms with $\lambda^n$, $n=1,2,\ldots,(k-1)$. We start with $k=N+1$ and 
construct algebraicaly independent invariants, step by step, by going to higher $k$.
 
For completeness, we give some results for $N=2$ case:
\baa
&&\Tr(A^{\dagger 2}A)
=\hlf\Tr A^{\dagger 2}\Tr A+\Tr A^{\dagger}\Tr(A^{\dagger}A)
-\hlf (\Tr A^{\dagger})^2\Tr A,\nn
&&\Tr(A^{\dagger 2}A^2)=\hlf\Tr A^{\dagger 2}\Tr A^2+
\Tr A^{\dagger}\Tr(A^{\dagger}A)\Tr A
-\hlf (\Tr A^{\dagger})^2(\Tr A)^2,\nn
&&:\Tr(A^{\dagger}AA^{\dagger}A):=-\hlf\Tr A^{\dagger 2}\Tr A^2+
:(\Tr(A^{\dagger}A))^2:+\hlf\Tr A^{\dagger 2}(\Tr A)^2+
\hlf(\Tr A^{\dagger})^2\Tr A^2\nn
&&-\hlf (\Tr A^{\dagger})^2(\Tr A)^2.\nonumber\ea

\end{document}